\documentclass[sigconf]{acmart}
\AtBeginDocument{%
  }

\copyrightyear{2024}
\acmYear{2024}
\setcopyright{rightsretained} 
\acmConference[CCS '24]{Proceedings of the 2024 ACM SIGSAC Conference on Computer and Communications Security}{October 14--18, 2024}{Salt Lake City, UT, USA}
\acmBooktitle{Proceedings of the 2024 ACM SIGSAC Conference on Computer and Communications Security (CCS '24), October 14--18, 2024, Salt Lake City, UT, USA}
\acmDOI{10.1145/3658644.3690298}
\acmISBN{979-8-4007-0636-3/24/10}

\makeatletter
\gdef\@copyrightpermission{
  \begin{minipage}{0.3\columnwidth}
   \href{https://creativecommons.org/licenses/by/4.0/}{\includegraphics[width=0.90\textwidth]{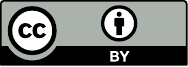}} 
  \end{minipage}\hfill
  \begin{minipage}{0.7\columnwidth}
   \href{https://creativecommons.org/licenses/by/4.0/}{This work is licensed under a Creative Commons Attribution International 4.0 License.}
  \end{minipage}
  \vspace{5pt}
}
\makeatother

\usepackage{subcaption}
\usepackage{soul}
\usepackage{url} 
\usepackage{graphicx}
\usepackage{xcolor}
\usepackage{enumitem}
\usepackage{amsmath}
\usepackage{multirow}
\usepackage{algorithm}
\usepackage{algpseudocode}

\DeclareMathOperator*{\argmin}{argmin}

\begin{document}

\title{\texttt{PromSec}: Prompt Optimization for Secure Generation of Functional Source Code with Large Language Models (LLMs)}

\author{Mahmoud Nazzal}
\email{mn69@njit.edu}
\orcid{0000-0003-3375-0310}
\affiliation{%
  \institution{New Jersey Institute of Technology}
  \city{Newark}
  \state{NJ}
  \country{USA}
}

\author{Issa Khalil}
\email{ikhalil@hbku.edu.qa}
\orcid{0000-0002-7660-9512}
\affiliation{%
  \institution{Qatar Computing Research Institute}
  \city{Doha}
  \country{Qatar}
}

\author{Abdallah Khreishah}
\email{abdallah@njit.edu}
\orcid{0000-0003-1583-713X}
\affiliation{%
  \institution{New Jersey Institute of Technology}
  \streetaddress{University Heights}
  \city{Newark}
  \state{NJ}
  \country{USA}
}
  
\author{NhatHai Phan}
\email{phan@njit.edu}
\orcid{0000-0002-1032-8275}
\affiliation{%
  \institution{New Jersey Institute of Technology}
  \streetaddress{University Heights}
  \city{Newark}
  \state{NJ}
  \country{USA}
}


\renewcommand{\shortauthors}{Khiem Ton, Nhi Nguyen, et al.}

\begin{abstract}
The capability of generating high-quality source code using large language models (LLMs) reduces software development time and costs. However, recent literature and our empirical investigation in this work show that while LLMs can generate functioning code, they inherently tend to introduce security vulnerabilities, limiting their potential. This problem is mainly due to their training on massive open-source corpora exhibiting insecure and inefficient programming practices. 
Therefore, automatic optimization of LLM prompts for generating secure and functioning code is a demanding need. This paper introduces \texttt{PromSec}, an algorithm for \underline{prom}pt optimization for \underline{sec}ure and functioning code generation using LLMs. 
In \texttt{PromSec}, we combine 1) code vulnerability clearing using a generative adversarial graph neural network, dubbed as gGAN, to fix and reduce security vulnerabilities in generated codes and 2) code generation using an LLM into an interactive loop, such that the outcome of the gGAN drives the LLM with enhanced prompts to generate secure codes while preserving their functionality. Introducing a new contrastive learning approach in gGAN, we formulate the code-clearing and generation loop as a dual-objective optimization problem, enabling \texttt{PromSec} to notably reduce the number of LLM inferences. As a result, \texttt{PromSec} becomes a cost-effective and practical solution for generating secure and functioning codes.

Extensive experiments conducted on Python and Java code datasets confirm that \texttt{PromSec} effectively enhances code security while upholding its intended functionality. Our experiments show that despite the comprehensive application of a state-of-the-art approach, it falls short in addressing all vulnerabilities within the code, whereas \texttt{PromSec} effectively resolves each of them. Moreover, \texttt{PromSec} achieves more than an order-of-magnitude reduction in operational time, number of LLM queries, and security analysis costs. Furthermore, prompts optimized with \texttt{PromSec} for a certain LLM are transferable to other LLMs across programming languages and generalizable to unseen vulnerabilities in training. 
This study presents an essential step towards improving the trustworthiness of LLMs for secure and functioning code generation, significantly enhancing their large-scale integration in real-world software code development practices.
\end{abstract}

\ccsdesc[500]{Security and privacy}
\ccsdesc[500]{Security and privacy~File system security}
\ccsdesc[500]{Security and privacy~Software security engineering}
\ccsdesc[500]{Security and privacy~Vulnerability scanners}
\ccsdesc[300]{Security and privacy~Formal security models}
\ccsdesc[300]{Security and privacy~Domain-specific security and privacy architectures}

\maketitle

\section{Introduction}
\par Large language models (LLMs) are at the forefront of machine learning (ML), demonstrating exceptional performance in various natural language processing (NLP) tasks. These models, including the generative pre-trained transformer (GPT) series by OpenAI \cite{radford2018improving,brown2020language}, utilize a large-scale, transformer-based architecture \cite{vaswani2017attention} trained on diverse NLP datasets. Recently, LLMs have attracted attention for generating high-quality source code, significantly benefiting software development by reducing time and expertise costs \cite{ziegler2022productivity}. Their growing popularity is evident, with tools like GitHub's Copilot attracting over a million paid subscribers across 37,000 organizations \cite{GitHubCopilot}. The effectiveness of these LLMs stems from their extensive training using vast source code databases. However, this reliance on open-source training data introduces significant security risks \cite{siddiq2022empirical}. The LLMs often replicate security flaws present in their training examples, leading to vulnerabilities in the generated code \cite{pearce2022asleep,henderson2018ethical,charalambous2023new,siddiq2023lightweight,perry2022users,sandoval2022security}. This issue raises concerns about the reliability of LLMs for secure code generation, also known as program synthesis, and highlights the necessity to address and mitigate these inherent security vulnerabilities effectively.

\par \textbf{Challenges.} The generation of secure and functionally correct code with LLMs faces the fundamental challenge of prompt optimization. Significant research efforts \cite{qin2021learning,zhou2022large, pryzant2023automatic,xu2022gps,li2023guiding} have been directed towards engineering and optimizing LLM prompts for specific applications, highlighting the vast prompt search space and developing heuristics to navigate this space effectively. Similarly, in code generation, a recent state-of-the-art (SoTA) approach \cite{pearce2023examining} shows the diversity in security and functionality of code generated with respect to variations in hand-crafted prompts. However, the search space challenge is more severe in the code generation application due to the need for external tools to assess code security and functional fidelity, making prompt optimization more complex. Moreover, the lack of differentiability of LLM prompts with respect to these security and functionality metrics, coupled with the absence of intuitive heuristics for maintaining security and functionality, adds to the challenge. Also, as detailed in Section \ref{Section4}, the one-to-many nature of function-code implementation further complicates the application of traditional, data-driven security solutions, particularly those based on a supervised learning approach. \texttt{PromSec} addresses these challenges using a graph generative adversarial network (gGAN) model, trainable on a differentiable contrastive loss that synergizes security reduction while upholding intended functionality. As elaborated in Section \ref{Section4}, this model effectively guides the LLM in generating better prompts that enhance security without compromising functionality. The model's effectiveness is demonstrated through an iterative interaction loop with the LLM, progressively fine-tuning the generation process to meet these dual objectives.

\par
\begin{figure}[!t]
\centering
\resizebox{0.99\columnwidth}{!}{
\includegraphics{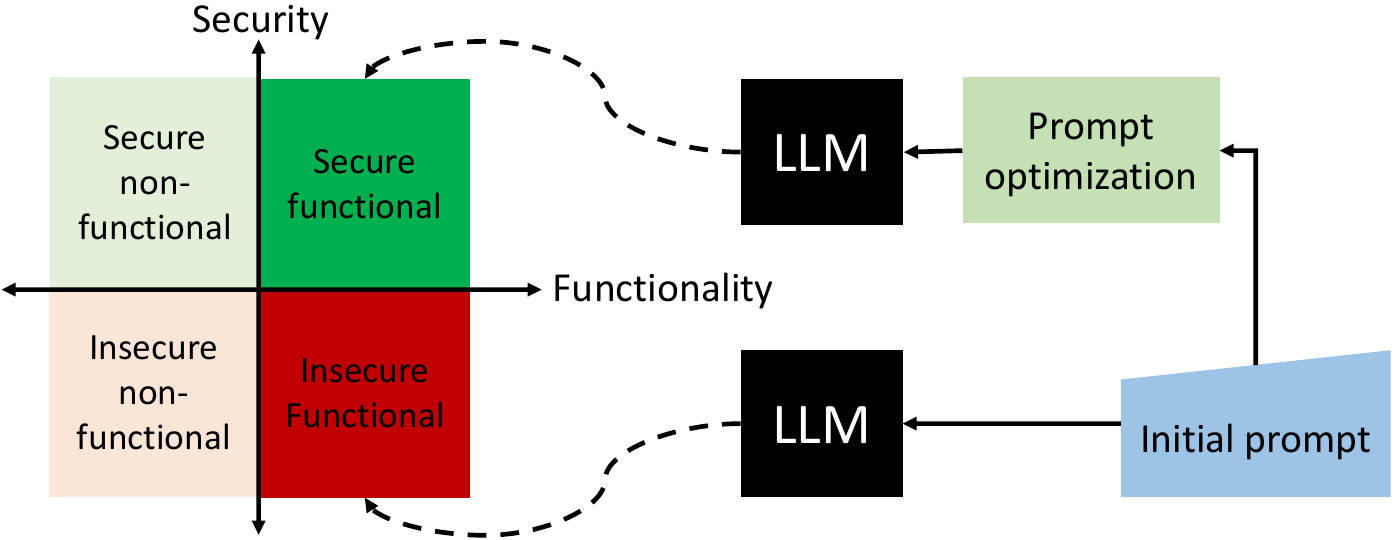}
}
\caption{With prompts from average users, LLMs tend to generate codes with security vulnerabilities.}
\label{basic_intro} 
\end{figure}

 \textbf{Prior Approaches and Limitations.} As pictorially depicted in Fig. \ref{basic_intro}, LLMs are inclined to produce functional and insecure code bases if one depends on regular prompts by average users. This is especially the case whenever the intended code's application has any security notation, as we empirically validate in Section \ref{Section3}. Efforts in prompt engineering for secure code generation, such as hand-crafted templates by Pearce et al. \cite{pearce2023examining}, are exhaustive and time-consuming. Methods like bounded model checking (BMC) \cite{charalambous2023new} are restricted to memory access issues. Other works \cite{islam2024code,islam2024llm} apply reinforcement learning to improve the security of LLM-generated code. However, these operate directly on source code rather than prompts. The only effort attempting prompt optimization is in \cite{he2023large}, but it requires white-box access, limiting its applicability to proprietary models like GPT and Bard. Overall, these approaches fall short of offering a unified framework for optimizing prompts for security and functionality, as depicted in Fig. \ref{basic_intro}. 


\par \textbf{\texttt{PromSec} Overview.} Based on the previous discussion, we formulate prompt optimization in code generation as a two-objective optimization problem aiming at code security and functionality preservation. To meet these objectives despite the non-differentiable nature of prompts with respect to their quantifications, we follow an iterative approach where we improve the code and then improve the prompt in each iteration. Code improvement is achieved by utilizing a gGAN model specially designed for this purpose; given an input code, it generates an output code with semantic similarity and reduced vulnerability. The model is trained to do so by minimizing a novel (differentiable) contrastive loss that involves the number of common weakness enumerations (CWE)\footnote{Common weakness enumerations (CWEs) are standardized categories that identify common software weaknesses and vulnerabilities. These are annually released by MITRE \cite{MITRE}, serving as a comprehensive reference to help organizations understand, identify, and address security issues in software.} as defined by MITRE \cite{MITRE} and the graph embedding distance. The LLM is then used to produce an improved prompt based on the improved code. In essence, the key difference this approach has over existing methods is the use of actual code fixes to steer the generation of improved prompts, rather than leaning on diagnostic tools such as what is done in \cite{pearce2023examining, charalambous2023new}. Our approach is anticipated to guide prompts toward a more thorough exploration of the solution space, as evidenced in performance assessments where vulnerabilities are swiftly resolved with minimal iterations at a low cost. Our focus is on fixing identified vulnerabilities, rather than vulnerability detection which is a distinct problem.

\par \textbf{Summary of Contributions.} The contributions in this paper are as follows.
\begin{itemize}[leftmargin=0pt]
 \item Introduction of \texttt{PromSec}, an algorithm for automatically optimizing prompts in LLM to generate secure source code while upholding its intended functionality.
 \item Development of a graph generative neural adversarial network (gGAN) model, characterizing semantic-preserving, and security-enforcing source code fixing as a two-objective optimization problem. This model, trained by minimizing a novel contrastive loss function, enables the application of semantic-preserving security fixes to the code graph.
 \item Demonstrating the efficacy of \texttt{PromSec} through comprehensive empirical validation, which confirms its capability in enhancing security and functionality in code generation using LLMs, and transferability across different LLMs, CWEs, and programming languages.
\end{itemize}

\par \textbf{Paper Outline.} Section \ref{Section2} presents relevant preliminaries. An empirical study of the need for code securing with LLMs is articulated in Section \ref{Section3}. Section \ref{Section4} details the proposed \texttt{PromSec} algorithm. Experiments and results are presented in Section \ref{Section5}. Section \ref{Section6} summarizes related work. A discussion is provided in Section \ref{Section7} with the conclusions in Section \ref{Section8}.

\section{Background}
\label{Section2}

\par \textbf{Code Generation with LLMs.} The idea of code generation traces back to sequence-to-sequence models, such as recurrent neural networks (RNN) and long short-term memory (LSTM). Here, the model takes a user's natural language prompt that represents their intent and produces a sequence of tokens representing the desired code. Subsequently, Transformer models \cite{vaswani2017attention}, and later LLMs \cite{radford2018improving,brown2020language, GPT35Turbo,CodeLlama,GoogleBard} have contributed to better code generation. Today, LLMs have gained prominence in this field. Trained on extensive datasets encompassing various programming languages and patterns, LLMs excel in producing syntactically correct, logically consistent, and functioning source code in many programming languages \cite{chen2021evaluating,austin2021program}. Their use extends beyond code generation to debugging, code translation, and even full software application development \cite{lu2021codexglue,chen2021evaluating}. As a result, employing LLMs can boost productivity by 30\% to 50\% in software development and reduce the costs of Information Technology (IT) by up to 65\% \cite{source55,source61,source62}. 
However, concerns remain about their ability to generate secure and vulnerability-free code consistently \cite{pearce2022asleep,henderson2018ethical,charalambous2023new,siddiq2023lightweight,perry2022users,sandoval2022security}.
The existence of such vulnerabilities in code can enable malicious exploitation, potentially leading to cyberattacks, data breaches, and system instability. Code vulnerabilities, including buffer overflow, injection attacks, and authentication bypass, can lead to unauthorized access and data theft \cite{comparitech2020, pentesttools2020}.

\par \textbf{Graph Representation of Source Code.} Graph representation of code abstracts relationships among different program elements and serves as input for a wide range of software analysis tasks \cite{wang2023graphspd,liu2023learning,allamanis2018learning,dinella2020hoppity,wang2020detecting}. The three widely used types of graph representations of code are abstract syntax trees (ASTs), control flow graphs (CFGs), and data flow graphs (DFGs) \cite{kavi1986formal}. Appendix \ref{appendixA}.1 illustrates constructing these graphs from a given code example. ASTs provide a hierarchical and structured representation of the syntactic elements within a program, enabling in-depth examination of code syntax and organization \cite{white2016deep}. CFGs, on the other hand, capture the control flow within a program, representing how control is transferred between different code blocks and aiding in understanding program execution paths \cite{zhao2018deepsim}. DFGs represent data dependencies and the flow of information within the program, offering insights into data-centric aspects of code behavior. 

\begin{table}[t]
 \centering
 \caption{Summary of major topics covered by our 50 prompts.}
 \label{PythonPromptsSummary}
 \resizebox{0.95\columnwidth}{!}{
 \begin{tabular}{|l|p{0.6\linewidth}|}
 \hline
 \textbf{Topic} & \textbf{Description} \\
 \hline
 Database Interaction & Interacting with databases, searching, deleting records, and data retrieval. \\
 \hline
 Security and Encryption & Data and file security, encryption, password hashing, and API key management. \\
 \hline
 Network and System Management & Network diagnostics, remote system monitoring, and IoT device control. \\
 \hline
 User Authentication & User login features and authentication with Python Flask. \\
 \hline
 File and Data Processing & Custom file managers, file operations, and user data processing. \\
 \hline
 Web Scraping and Web Apps & Web scraping, rendering HTML content, and automating web-related tasks. \\
 \hline
 Mathematical and Data Analysis & Calculators, data processing tools, and mathematical operations. \\
 \hline
 \end{tabular}}
\end{table}

Graph-based code repair has emerged as a useful tool mainly applied for detecting malfunctioning portions of source code (patches) and suggesting remedies to them. As leading graph representation models, GNNs can effectively capture the complex relationships and dependencies inherent in programming languages \cite{allamanis2018learning,dinella2020hoppity}. In the context of automatic code repair, GNNs are mainly used to identify buggy portions of source code thereby hinting at potential fixes \cite{tufano2019empirical,chen2019sequencer}. These fixes can range from minor syntax corrections to significant structural changes aimed at closing security gaps.

\par \textbf{GNNs and Generative GNNs.} Graph neural networks (GNNs) extend deep learning to graph data, utilizing neural network layers for message passing and aggregation \cite{welling2016semi}. Efficiently transforming graph information into node embeddings, GNNs excel in capturing complex relationships within graphs, leading to SoTA performance in various applications \cite{wu2020comprehensive}. Among generative GNNs, graph generative adversarial networks (gGANs) and graph variational autoencoders (gVAEs) are particularly noteworthy. A gGAN \cite{wang2018graphgan} employing adversarial frameworks consists of a \textit{generator} and a \textit{discriminator}, for complex network synthesis. The generator creates graphs, while the discriminator assesses their authenticity. Conversely, a gVAE \cite{kipf2016variational} uses a variational autoencoder architecture, encoding graph data into a latent space and then decoding it to generate new graphs. 

\par \textbf{Contrastive Learning.} Contrastive learning is a model training technique that emerged as a remedy for the data-labeling requirement of supervised learning. Instead of using specific data labels, a contrastive learning scheme identifies positive and negative data pairs. A positive pair comprises two data points of relatively strong similarity as opposed to a negative pair. Therefore, the model is trained to minimize the differences between items in a positive pair and maximize it for items in a negative pair. A clear advantage of this learning scheme is alleviating the need for data labels explicit supervision \cite{chen2020simple,he2020momentum}. Contrastive learning has shown notable advantages in tasks like image and text representation \cite{chen2020simple}, and its applicability extends to the domain of GNNs. In GNNs, specifically generative GNNs, contrastive learning facilitates the learning of node and graph embeddings that accurately capture the underlying structure and features of graph data \cite{velivckovic2018deep,you2020graph,wang2020detecting}. As a result, contrastive learning and generative GNNs are particularly suitable for tasks where understanding the relationships and structures within data is crucial, such as in social network analysis, recommendation systems, and bioinformatics.

\begin{figure}[htb]
\centering
 \resizebox{0.99\columnwidth}{!}{
\begin{tabular}{cc}
\includegraphics[width=8cm]{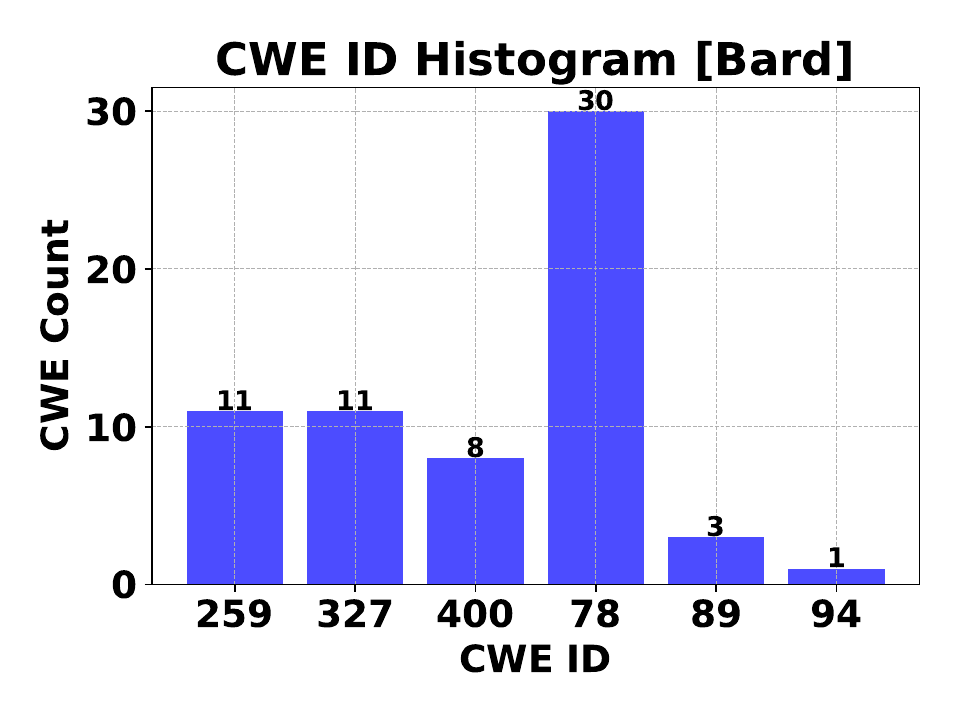}
&
\includegraphics[width=8cm]{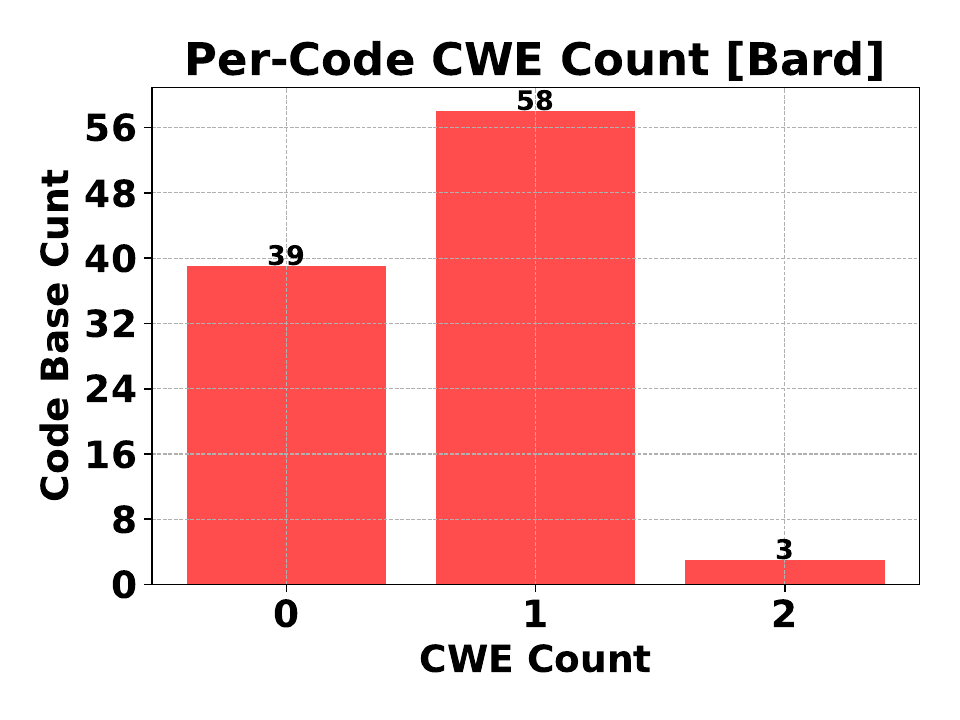}
\\
\Huge{(a)} & \Huge{(b)} 
\\
\includegraphics[width=8cm]{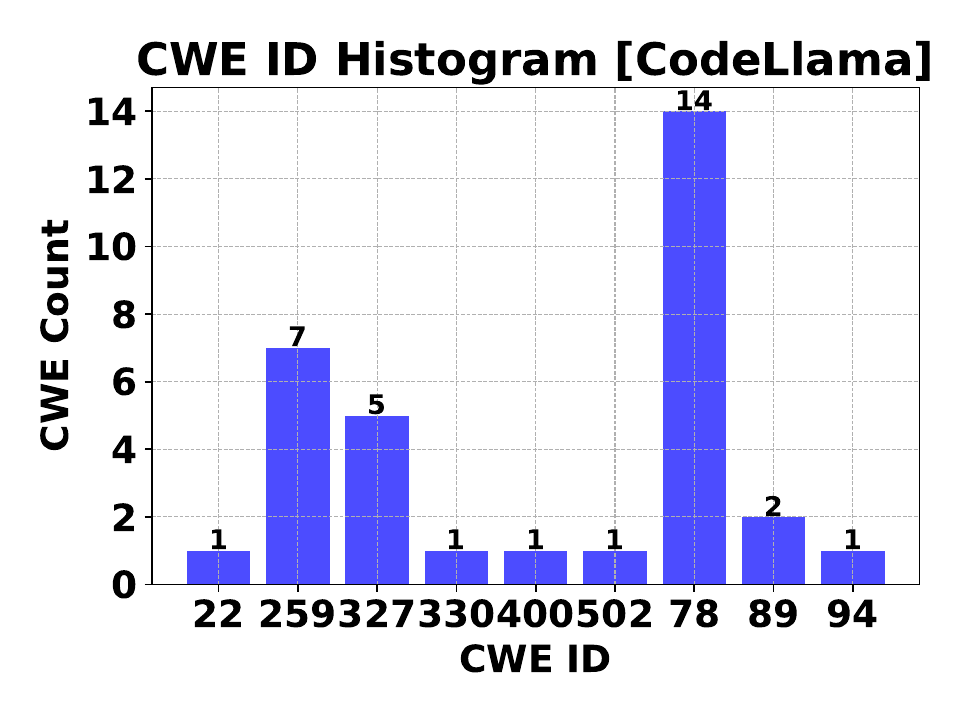}
&
\includegraphics[width=8cm]{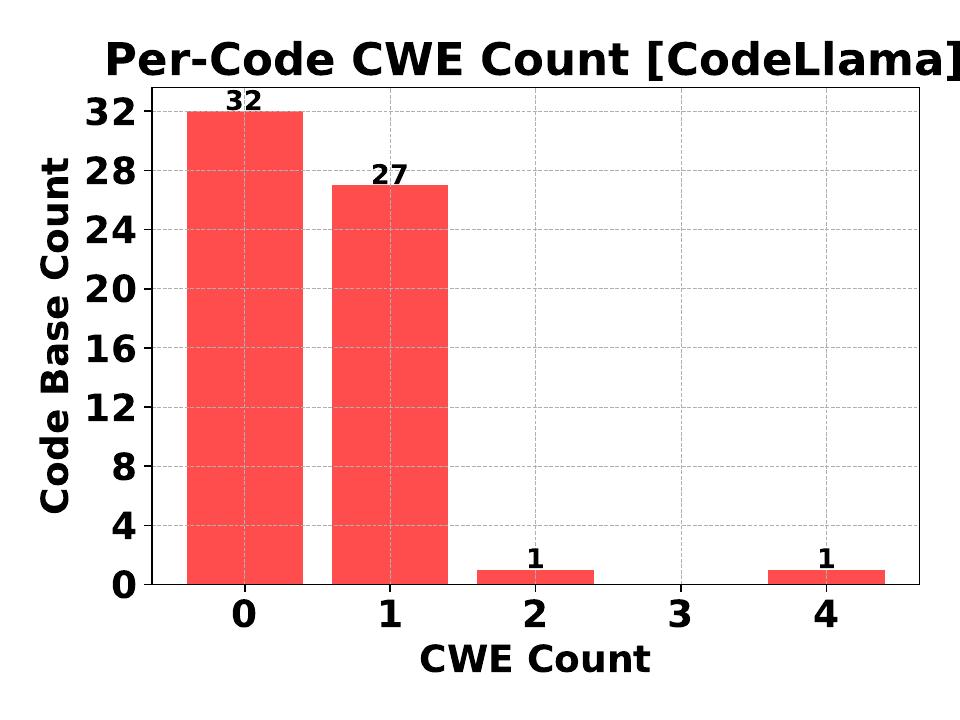}
\\
\Huge{(c)} & \Huge{(d)} 
\\
\includegraphics[width=8cm]{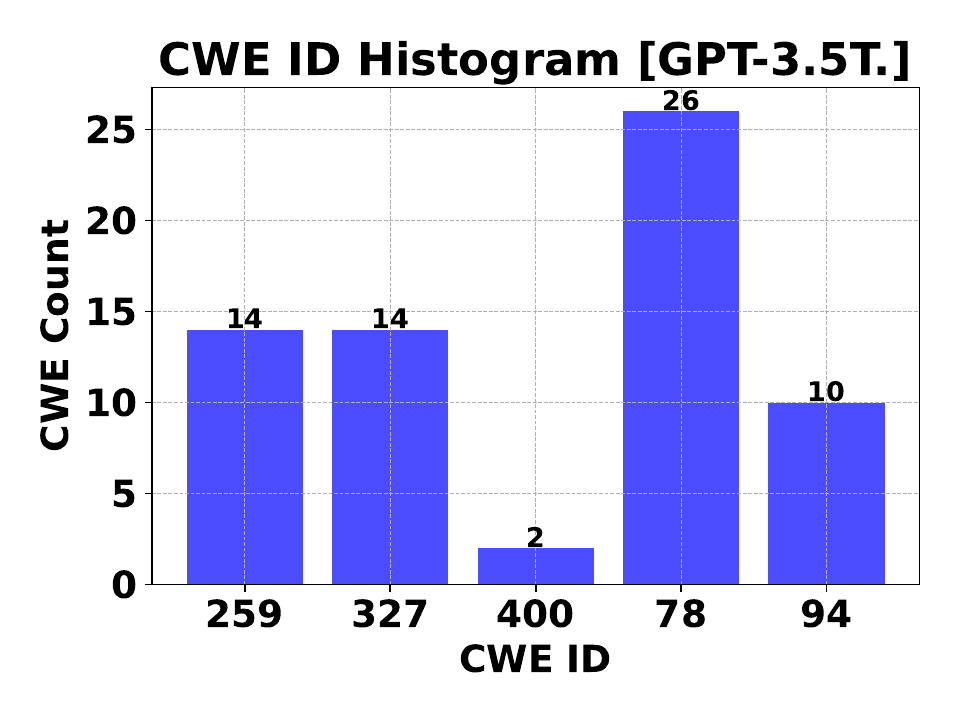}
&
\includegraphics[width=8cm]{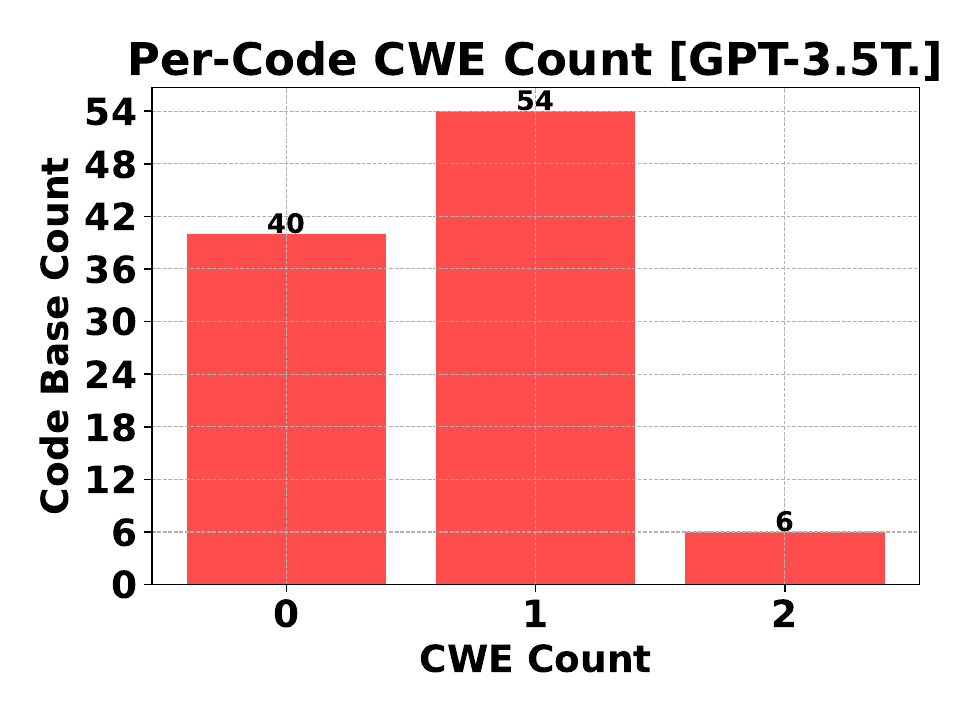}
\\
\Huge{(e)} & \Huge{(f)} \\
\includegraphics[width=8cm]{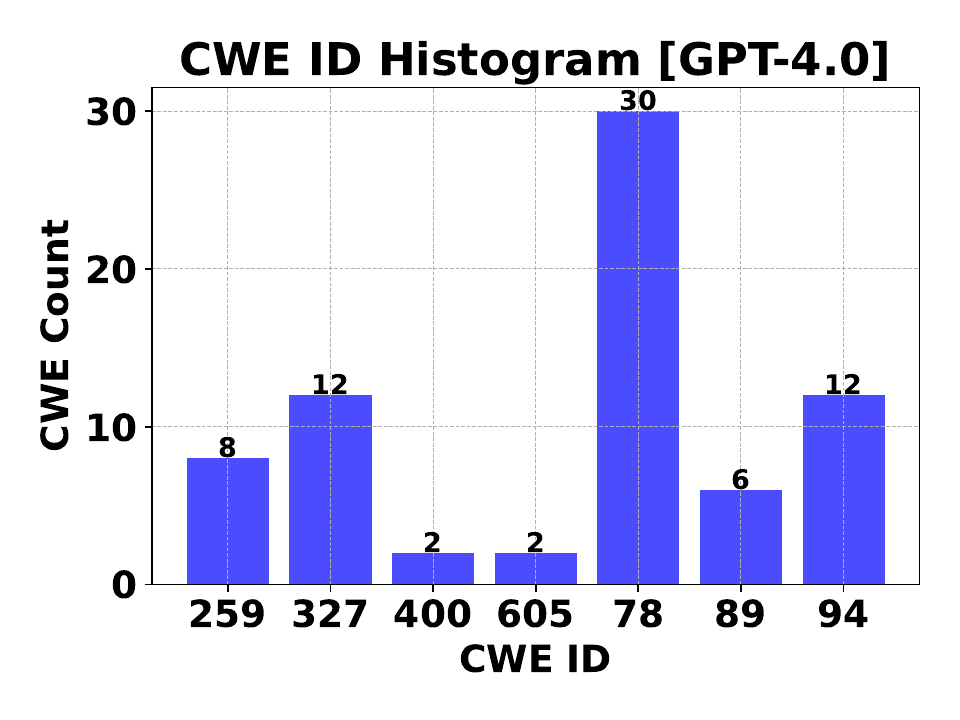}
&
\includegraphics[width=8cm]{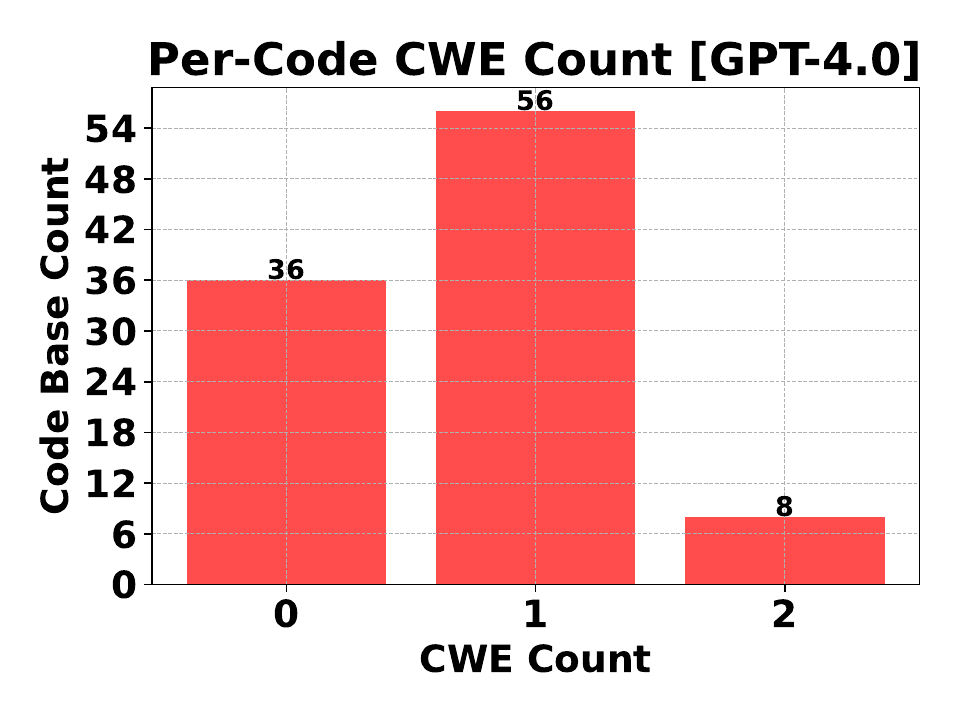}
\\
\Huge{(g)} & \Huge{(h)} \\
\end{tabular}}
\vspace{-0.7em}
\caption{Histogram of CWEs in 100 generated code bases according to the prompts summarized in Table \ref{PythonPromptsSummary} (left) and the histogram of the CWE count per code base (right), for Bard \cite{GoogleBard}, CodeLlama-70B-Instruct \cite{CodeLlama70binstrcut}, GPT-3.5 Turbo \cite{GPT35Turbo}, and GPT4 \cite{GPT4} row-wise, respectively.}
\label{empir_moativ1} 
\end{figure}

\section{How Secure are LLMs in Code Generation?}
\label{Section3}
\par Since LLMs are commonly perceived to generate code bases prone to security vulnerabilities \cite{pearce2022asleep,henderson2018ethical,charalambous2023new,siddiq2023lightweight,perry2022users,sandoval2022security}, our initial focus is to assess this concern on modern LLMs. To achieve this objective, we conducted the following experiment. We manually design 50 prompts, each asking the LLM to generate a code base that performs a specific task. While tasks have some security notions, these prompts do not explicitly ask the LLMs to beware of security issues, nor do they represent bad programming practices. They are just subtle verbal descriptions of what an LLM is requested to generate, and they are aimed to represent the skill level of an ordinary LLM user without security expertise. Table \ref{PythonPromptsSummary} summarizes the main topics covered by these prompts. These topics cover a wide range of Python scripting and application development areas, including database access, security, encryption, network and system management, user authentication, file and data processing, web scraping, and mathematical and data analysis.

\par We feed the prompts to a set of prominent LLMs including Google's Bard \cite{GoogleBard}, Meta's CodeLlama-70B-Instruct \cite{CodeLlama70binstrcut}, OpenAI's GPT-3.5 Turbo \cite{GPT35Turbo}, and OpenAI's GPT4 \cite{GPT4}, and observe their generated codes. Then, we analyze the security of the generated code bases using widely applied tools such as Bandit \cite{bandit_ref} and SpotBugs \cite{spotbugs2023}. Each prompt is fed to each LLM twice, resulting in 100 generated code bases per LLM. Fig. \ref{empir_moativ1} shows the outcomes of the security analysis across these LLMs presented in rows, respectively. For each LLM (row), we first plot a histogram of the identified CWEs in the generated codes (left) and a histogram of the counts of CWEs in each code base (right). The figure supports previous studies \cite{pearce2022asleep,henderson2018ethical,charalambous2023new,siddiq2023lightweight,perry2022users,sandoval2022security} that LLMs generate codes prone to security vulnerabilities as evidenced by the identified CWEs. 
\par We note that CWE 78 stands out as the most frequently recurring security vulnerability among the identified ones. There is a noticeable overlap in the identified CWEs across the four LLMs. Moreover, with Bard, GPT-3.5 Turbo, and GPT-4.0, the security analysis reveals that no more than 40\% of the generated code bases are entirely secure, according to the security analysis tools we use for Python and Java codes. It is noted that CodeLlama-70B-Instruct generates code bases in 61 of the times it is prompted. This is consistent with the common belief that Llama models are overly protective and tend to have a high \textit{false rejection rate} \cite{bhatt2024cyberseceval}. However, the code bases generated still have similar vulnerability occurrences. We emphasize that these findings are derived without prompting the LLMs to explicitly or implicitly create CWEs. CWEs surfaced solely because the LLMs were instructed to generate code for security-related tasks. This experiment illustrates that SoTA LLMs often produce insecure code when tackling security-related tasks. Hence, it is crucial not to unquestionably rely on their outputs. 

\begin{figure}[!thb]
\centering
\resizebox{0.86\columnwidth}{!}{
\includegraphics[width=8cm]{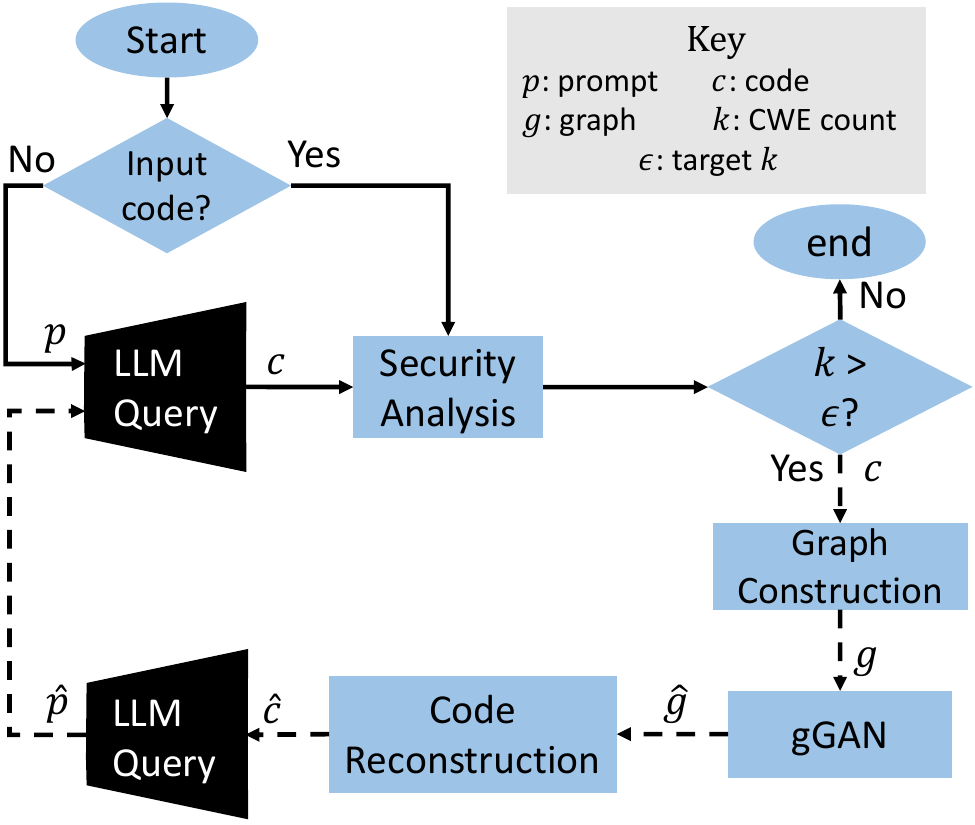}
}
\caption{An overview of \texttt{PromSec}'s pipeline.}
\label{overall_system} 
\end{figure}

\section{\texttt{PromSec}}
\label{Section4}

\par The key idea behind \texttt{PromSec} is applying contrastive learning within a graph generative adversarial network (gGAN) combined with static security analysis. This unified method aims to direct the LLM in producing secure and functioning code as an alternative to conventional optimization approaches that are hindered by the non-differentiable nature of generating secure and functioning code.

\par Fig. \ref{overall_system} shows the main steps of our proposed \texttt{PromSec} approach. First, an initial prompt is obtained. Such a prompt is taken directly from the user as a verbal description characterizing her/his intent \cite{le2020deep}. The initial prompt is forwarded to the LLM and a code is generated. The generated code is then forwarded to the \textit{Security Analysis} module, which is a static security tool to identify the number of CWEs in the code. Alternatively, if the input is source code $c$, it is passed directly to the \textit{Security Analysis} module. The code is considered the final output if the security analysis identifies no CWEs, denoted as ``$k=0$'' (where `$k$' represents the count of identified CWEs by the security analysis tool). Otherwise, the code is transformed into a graph representation $g$ (e.g., AST) using the \textit{Graph Construction} module, then forwarded to a \textit{gGAN}. This gGAN model aims to enhance the input graph representation, striving for a more secure (with reduced CWEs) one, $\hat{g}$, while maintaining the intended functionality. Once generated, the graph $\hat{g}$ is converted back into code $\hat{c}$ through a \textit{Code Reconstruction} module. This reconstructed code is then reverse-engineered by the LLM\footnote{By explicitly instructing the LLM to analyze the codebase and estimate a detailed prompt that could have generated it.} to generate a prompt corresponding to $\hat{c}$, $\hat{p}$. The new prompt $\hat{p}$ is fed back into the LLM, generating new code versions in a loop that continues until the CWE count reduces to a small desired value ($\epsilon$), which ideally is zero) or for a prescribed number of cycles.

\par In this section, we first establish the suitability of a contrastive learning framework for the training of the aforementioned gGAN model. Then, we provide the architectural details and the proposed contrastive loss to be used in the training of this model along with its training algorithm.


\subsection{Code Fixes are One-to-Many}

\par Function-implementation relationship in code bases is well-known to be a one-to-many mapping, i.e., the same semantic function can be implemented as a code in many ways \cite{lee2021toward,trisovic2022large}. Furthermore, code similarity detection systems like Facebook's Aroma and Intel's Machine Inferred Code Semantics Analysis (MISIM) illustrate the ability to recognize different code implementations fulfilling identical functions, supporting the one-to-many mapping \cite{lee2021toward}.

\par It is interesting to discuss the implication of the one-to-many function-to-code relationship on suitable learning of generative models on code graphs. Let us recall that a code can be represented as a graph. Thus, it is possible to obtain an improved code by utilizing a generative GNN model trained to generate better code graphs. In supervised learning of generative models, model training requires labeled data, where a data point label represents the desired outcome that the model needs to generate. In an application like source code fixing, an immediate approach is to simply feed a generative model with an input code graph and train it to generate a secure version of this code graph. However, recalling the aforementioned one-to-many function-code mapping, it is evident that the same code can have many secure versions. In other words, for a given insecure code snippet, there could be multiple and structurally diverse but functionally equivalent secure versions. Accordingly, there is no notion of a single desired outcome or, equivalently, no notion of a unique data label.

To demonstrate this limitation, we conduct the following experiment. We consider a set of 100 Python codebases, each containing multiple CWEs. For each codebase, we generate three CWE-free \textit{versions}. Specifically, each test codebase is provided to the LLM, which is instructed to generate a secure version that retains the same functionality. This secure version is analyzed using a security tool to ensure it is free of CWEs and is manually verified to maintain the same functionality. This process repeats until three secure versions are obtained per sample. Next, we define and calculate the average \textit{inter-version} distance, which is the mean graph edit distance between the code graph of the original test codebase and that of its corresponding CWE-free versions. We also calculate the average \textit{intra-version} distance, which denotes the average graph edit distance among the graph representations of the CWE-free versions of each codebase. We calculate these metrics for the 100 considered codebases.

\begin{figure}[htb]
\centering
\resizebox{0.84\columnwidth}{!}{
\includegraphics{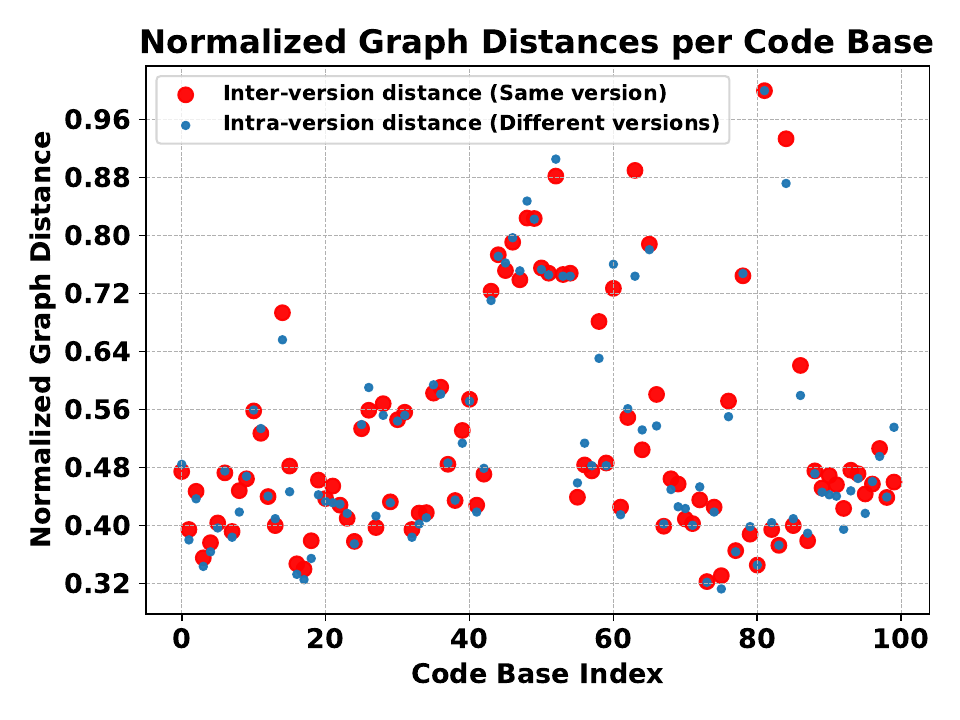}
}
\vspace{-0.9em}
\caption{A comparison of the average inter-version and average intra-version graph edit distances per code base.}
\label{motiv2} 
\end{figure}

\par Fig. \ref{motiv2} illustrates the results of those two metrics per code base (normalized by the maximum value). It is worth noting that the average inter-version and average intra-version distances are comparable. Furthermore, on many occasions, the inter-version distance is even less than the intra-version distance.
This observation indicates that the edit distance between the original code base and its CWE-free versions is less than or at least comparable to the distance among its CWE-free versions.

\par The observed diversity in secure and functioning code versions suggests the viability of integrating contrastive learning in the gGAN model for code improvement alongside static security analysis. This combination provides an effective alternative approach for guiding LLMs in generating secure and functioning code using optimized prompts. Contrastive learning's strength lies in distinguishing between various examples, making it suitable for scenarios with multiple valid outcomes. It can train models to recognize a broad range of secure coding patterns beyond replicating a single secure version, expanding their understanding of diverse secure coding practices. 

\subsection{The Training and Operation Pipeline}

\par We formulate prompt optimization for secure and functioning code generation with LLMs as follows: 
\begin{equation}
p^* = \argmin_{p} \big[ \alpha k(c) + \beta d(c, c_0) \big],
\label{eq1}
\end{equation}
\noindent where $p$ is a prompt, $p^*$ is the optimal prompt, $c$ is the code to be generated, $c_0$ is the original code either given as a code starting point or the code generated with an initial prompt $p_0$. The term $k(c)$ denotes the CWE count in the code $c$, and $d(c, c_0)$ is a metric representing the functional discrepancy between $c_0$ and $c$. Also, $\alpha$ and $\beta$ are weight parameters.

\par As highlighted in the Introduction Section, the essential challenge in this optimization is that $p$ is not differentiable with respect to $k(c)$ and $d(c, c_0)$. Even modeling these functions in terms of $p$ is challenging since it requires modeling the LLM with external security analysis tools. 
To address this challenge, we approximate the optimal prompt $p^*$ in Eq. \ref{eq1} by using a gGAN model to generate improved code graphs that lead to improved prompts for generating secure and functioning code in an iterative manner, as follows.

\par gGAN rectifies code issues by altering its graph representations, which are then translated into prompt adjustments. Operating on graph data, this model follows the framework akin to the generator-discriminator setup in classical GAN networks. Here, inputs drive the generator to produce outputs meeting specific criteria, while the discriminator evaluates the adherence of these outputs to the defined requirements. As it operates on graph data, it is classified as a GNN-based GAN. Architectures such as the graph convolutional network (GCN)-based GAN \cite{lei2019gcn} can be utilized in this context. 

\par The proposed learning mechanism for the gGAN is outlined in Algorithm \ref{algorithm1}. We train the model using a unique contrastive loss for the generator, ensuring the generation of secure and functioning code graphs. We formulate the contrastive loss of the gGAN as follows:

\par The \textbf{generator loss} is composed of adversarial and contrastive losses, denoted as $\mathcal{L}_{\text{adv}}$ and $\mathcal{L}_{\text{contrastive}}$ respectively, balanced by a hyperparameter $\lambda$ as
\begin{equation}
\mathcal{L}_G = \mathcal{L}_{\text{contrastive}} + \lambda \mathcal{L}_{\text{adv}}.
\label{eq2}
\end{equation} 

\begin{itemize}[leftmargin=0pt]
 \item \textbf{Adversarial Loss \cite{goodfellow2014generative}.} The adversarial loss $\mathcal{L}_{\text{adv}}$ evaluates how effectively the generator $G$ can deceive the discriminator $D$. It is defined as 
\begin{equation}
 \mathcal{L}_{\text{adv}} = -\mathbb{E}[\log D(G(g_i))]. 
 \label{eq3}
\end{equation}

\item \par \textbf{Contrastive Loss.} The contrastive loss $\mathcal{L}_{\text{contrastive}}$ is incorporated in terms of the number of CWEs (denoted by $k$) and an embedding similairty measure $S$:
\end{itemize}
\begin{align}
& \mathcal{L}_{\text{contrastive}} = -\log (x), \label{eq4} \\
& \text{s.t. } x= \frac{e(\alpha \Delta k_i + \beta S_i)}{e(\alpha\Delta k_i + \beta S_i) + \sum_{j \neq i} e(\alpha\Delta k_{ij} + \beta S_{ij})},
\label{eq5}
\end{align}
\noindent where $\Delta k_i = k(c_i) - k(\hat{c}_i)$ is the difference in CWE counts between the code $c_i$ represented by an input graph $g_i$ and its fixed counterpart $\hat{c}_i$ represented by the generated graph $\hat{g}_i$, $S_i$ is a graph embedding similarity measure between the two graphs $g_i$ and $\hat{g}_i$, $\Delta k_{ij}$ is the difference in CWE counts of the code bases represented by graphs $g_i$ and $g_j$, $S_{ij}$ is their graph similarity, and $e(z) \triangleq \exp(-z)$, $\alpha$ and $\beta$ are used to control balancing code security and functional similarity, and the summation is over graphs in the training dataset.

\par The design of the contrastive loss function aims to encourage the generation of graphs with fewer CWEs and greater similarity in embeddings between the original and generated graphs, thereby maintaining code functionality \cite{zhao2018deepsim,fang2020functional,wang2020detecting,liu2023learning}.

\par \textbf{Discriminator Loss.} The discriminator loss $\mathcal{L}_D$ follows the standard GAN format, being the binary cross-entropy loss between real and generated graphs \cite{goodfellow2014generative}, as follows:
\begin{equation}
\mathcal{L}_D = -\mathbb{E}[\log D(g_i)] - \mathbb{E}[\log(1 - D(G(g_i)))].
\label{eq6}
\end{equation}

\par \textbf{Gradient Updates.} For the discriminator and generator modules, gradients of these loss functions are computed and used for updating the respective parameters through stochastic gradient descent (SGD) \cite{robbins1951stochastic}, as follows: 
\begin{equation}
\theta_D \leftarrow \theta_D - \eta \nabla_{\theta_D} \mathcal{L}_D,
\label{eq7}
\end{equation}
\begin{equation}
\theta_G \leftarrow \theta_G - \eta \nabla_{\theta_G} \mathcal{L}_G,
\label{eq8}
\end{equation}
where \( \eta \) represents the learning rate.

\begin{algorithm}
\caption{Training of the gGAN Model}
\small
\begin{algorithmic}[1]
\State \textbf{Input:} Training graphs $g_1, g_2, \ldots, g_n$, number of iterations $T$
\State \textbf{Initialize:} generator parameters $\theta_G$ and discriminator parameters $\theta_D$
\For{$t=1$ to $T$}
 \For{$i=1$ to $n$}
 \State Sample a graph $g_i$
 \State Generate an improved graph $\hat{g}_i$ using $\theta_G$
 \State Update $\theta_D$ according to Equation (\ref{eq7})
 \State Update $\theta_G$ according to Equation (\ref{eq8})
 \EndFor
\EndFor
\State \textbf{Output:} Trained generator and discriminator ($\theta_G$ and $\theta_D$)
\end{algorithmic}
\label{algorithm1}
\end{algorithm}

\par \textbf{Code-to-Graph Conversion.} In the proposed pipeline, code can be easily converted into a graph representation (e.g., AST, CFG, DFG) using off-the-shelf parser packages such as \cite{Python39Parser,AstUnparseDocs}. Similarly, converting an AST graph into code is a straightforward task, achieved by unparsing the AST. However, converting a CFG to code is more complicated.

\begin{figure}[t]
\centering
\resizebox{0.92\columnwidth}{!}{
\includegraphics{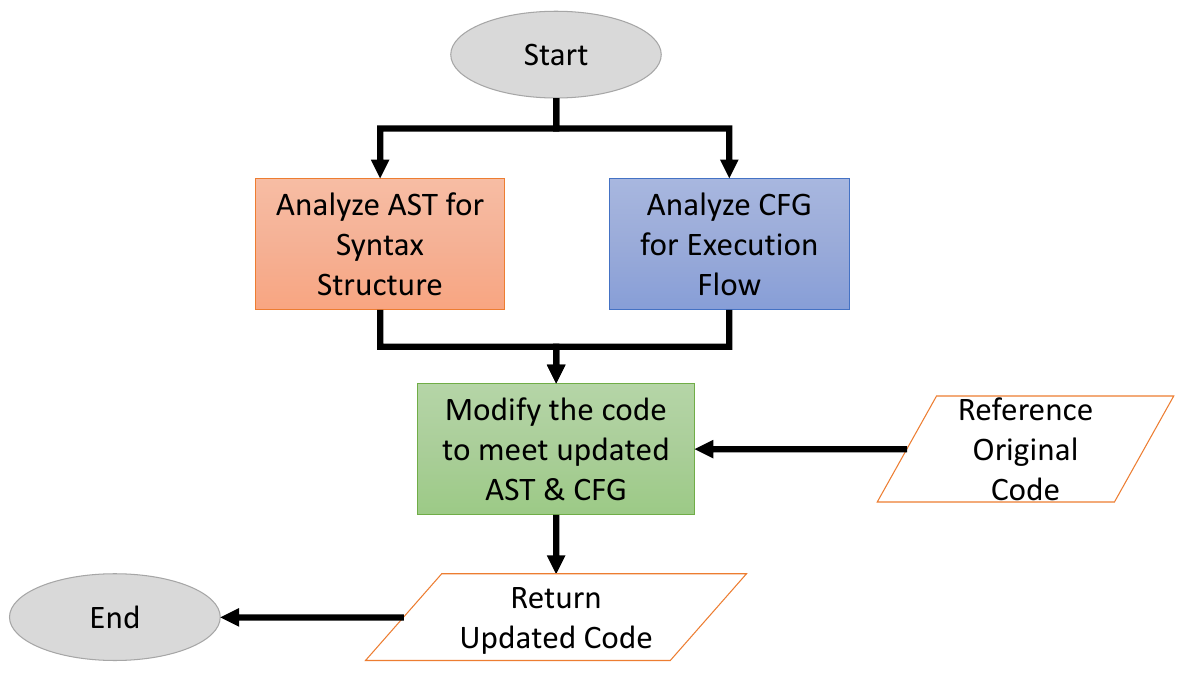}
}
\caption{Code reconstruction from CFG graph edits.}
\label{CodeReconstructionProcess} 
\end{figure}

To overcome this challenge, we employ a method that considers the initial code, its AST, and CFG graphs alongside the new CFG to reconstruct the updated code as illustrated by the flowchart in Fig. \ref{CodeReconstructionProcess}. This approach involves modifying elements of the original code based on discrepancies between the CFG and AST graphs of the original and updated versions. These modifications are guided by the need for the new code to align with the new CFG while maintaining consistency. This strategy ensures the integration of the new code changes while upholding the foundational style and structure of the original code. DFG to code conversion follows a process similar to that of CFG.

\section{Experiments}
\label{Section5}

\par We present here the results of the experiments conducted to evaluate \texttt{PromSec}'s performance. In summary, the results validate that \texttt{PromSec} outperforms existing baselines in terms of security, functionality preservation, and various aspects of cost-effectiveness. Importantly, prompt optimizations obtained with \texttt{PromSec} are generalizable to unforeseen CWEs, across different types of LLMs, and different programming languages. The source code and data are available at the link: \href{https://github.com/mahmoudkanazzal/PromSec}{https://github.com/mahmoudkanazzal/PromSec}. In the following subsection, we aim to answer the research questions listed in Table~\ref{summarytable}.

\begin{table}[t]
\centering
\caption{A summary of research questions and answers.}
\resizebox{0.99\columnwidth}{!}{
\begin{tabular}{|c|l|l|}
\hline
Q & Summary & Key Result \\ \hline
1 & How successful is \texttt{PromSec} in enforcing code security? & Highly \\ \hline
2 & Does \texttt{PromSec} need iterative LLM interaction? & Yes \\ \hline
3 & How does \texttt{PromSec} compare to SoTA baselines? & Superior \\ \hline
4 & How successful is \texttt{PromSec} in functionality preservation? & Highly \\ \hline
5 & What is the impact of code graph types on \texttt{PromSec}? & CFGs perform best \\ \hline
6 & Do \texttt{PromSec} prompts transfer to new CWEs? & Yes \\ \hline
7 & How is \texttt{PromSec}'s transferability across LLMs? & High\\ \hline
8 & How is \texttt{PromSec}'s transferability across programming languages? & High\\ \hline
\end{tabular}
}
\label{summarytable}
\end{table}

\begin{figure}[!tb]
\centering
\resizebox{0.99\columnwidth}{!}{
\begin{tabular}{c}
\includegraphics[width=15cm]{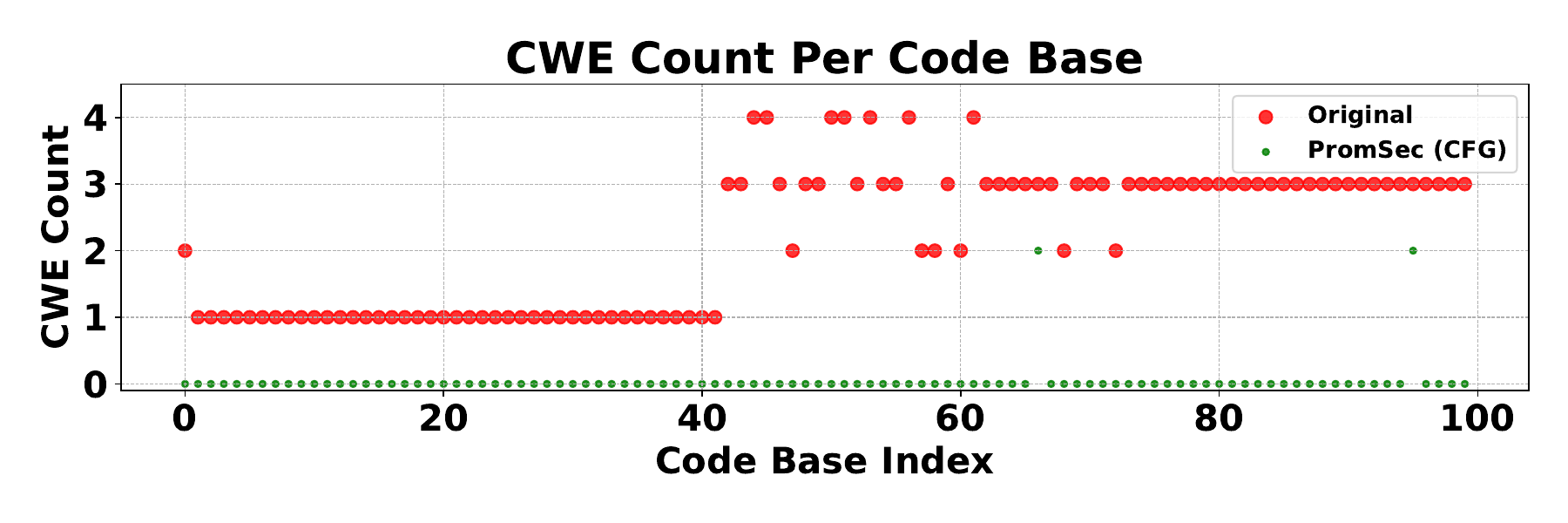}
\\
\Huge{(a)}
\\
\includegraphics[width=15cm]{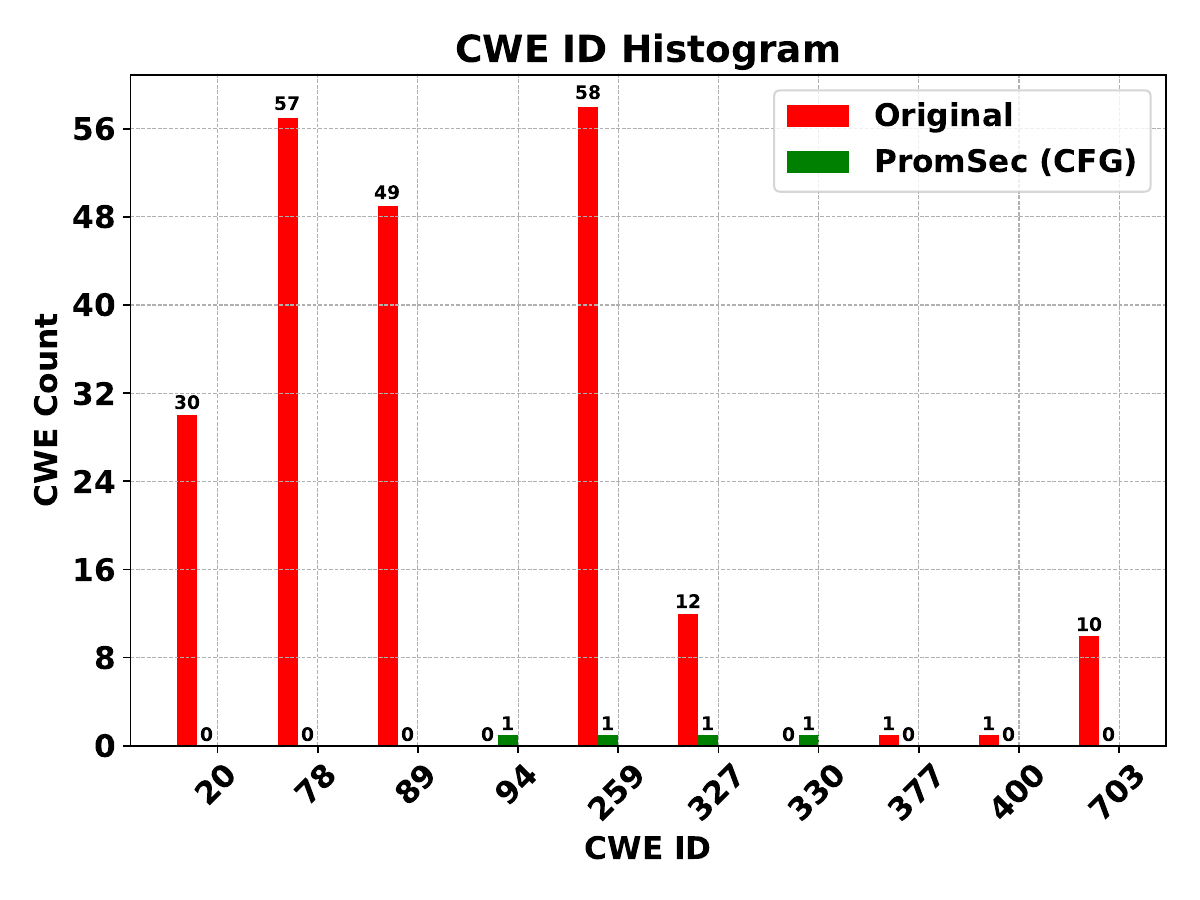}
\\
\Huge{(b)}
\vspace{-1em}
\end{tabular}}
\caption{Performance evaluation plots with CFG-type graphs: (a) per-code base CWE count and (b) the histogram CWEs before and after applying \texttt{PromSec}.}
\label{pa_cfg} 
\end{figure}

\subsection{The Setup, Dataset, and Baselines}

\par The setup includes an LLM used for code generation. An LLM prompt is a verbal description of the intent for generating the code. This description is either obtained directly from an assumed user or by instructing the LLM to generate an approximated prompt in case of starting with source code, as detailed in Section~\ref{overall_system}. Unless stated otherwise, the default LLM used is GPT-3.5 Turbo. 
The gGAN model used has a Generator with two GraphConv layers and ReLU activations, and a Discriminator with two GraphConv layers with ReLU and Sigmoid activations, implemented using PyTorch Geometric. In our experiments, we set $\alpha$ and $\beta$ appearing in equations (\ref{eq1}) and (\ref{eq5}) to 1. We found that increasing $\alpha$ over $\beta$ slightly enhances security but reduces code functional similarity, and vice-versa. We use two datasets. The first is a Python code base set obtained from \cite{pearce2023examining} and comprising source codes from MITRE's \cite{MITRE} documentation, CodeQL documentation \cite{GitHubInc2021CodeQL}, and hand-crafted codes by the authors of \cite{pearce2023examining}. We refer the reader to Appendix \ref{appendixA}.2 for more details on this dataset. The second is a dataset of Java prompts collected from \cite{hao2022aixbench,yu2023codereval}. For the security analysis, we select Bandit \cite{bandit_ref} for Python and SpotBugs \cite{spotbugs2023} for Java, for their wide popularity and outstanding performance in terms of detection quality and time complexity. However, any other static security tools can be used for this purpose.

\begin{table}[t]
\centering
\small
\caption{Context degree across prompt templates in BLs.}
\vspace{-0.7em}
\resizebox{0.75\columnwidth}{!}{
 \begin{tabular}{|c|l|}
 \hline
 Template & Security Context \\
 \hline
 1 & None (No help) \\\hline
 2 & CWE line numbers \\\hline
 3 & CWEs IDs \\\hline
 4 & CWE line numbers, CWEs IDs \\\hline
 5 & CWE line numbers, confidence \\\hline
 6 & CWEs IDs, confidence \\\hline
 7 & CWE line numbers, CWE IDs, confidence \\
 \hline
 \end{tabular}
}
\label{template_table}
\end{table}

\par We compare the performance of \texttt{PromSec} with the following baselines (BLs):
\begin{itemize}[leftmargin=0pt]
 \item \textbf{BL1} characterizes Pearce et al.'s work \cite{pearce2023examining} accustomed to code generation. We define 7 prompt templates with increasing context. The contexts of templates are summarized in Table \ref{template_table}. It is because security analysis tools typically return the IDs of identified CWEs, the lines of code where they occur, and the confidence in their detection.
 \item \textbf{BL2} applies BL1 iteratively and at each iteration, greedily selecting the solution with the minimal CWEs count.
\end{itemize}

\par We consider the following metrics in our experiments:
\begin{enumerate}[leftmargin=0pt]
 \item \textbf{Security Enhancement.} Quantified by the reduction in the number of CWEs ($k$) as commonly employed in \cite{pearce2022asleep,pearce2023examining,siddiq2023lightweight,siddiq2023generate,he2023large}. This metric is computed based on the outcomes of static security analysis techniques that detect and enumerate CWEs.
 
 \item \textbf{Preservation of Code Functionality.} Quantified by the similarity between code graphs as commonly employed in many works such as \cite{zhao2018deepsim,fang2020functional,wang2020detecting,liu2023learning}. To supplement this, we conduct fuzzing tests \cite{li2018fuzzing} on a randomly selected subset of generated code bases, providing an additional layer of evaluation. 
 
 \item \textbf{Operational Cost.} Evaluated based on three parameters, including the time taken for code execution (execution time delay), the financial and delay implications of utilizing LLMs, and the costs of static security analysis.
\end{enumerate}

\subsection{Performance Evaluation of \texttt{PromSec}}
In this subsection, we answer Q1 through Q4 (Table \ref{summarytable}) considering CFG as the graph representation. We examine the other two code graph types (AST, DFG) in Subsection \ref{Section5}.3.

\par First, we address \textbf{Q1: How successful is \texttt{PromSec} in code security enforcement?}
We train the gGAN model with 500 training graphs. Then, we feed 100 test code bases to \texttt{PromSec} and set the maximum number of iterations to 20. First, with a CFG graph type, Fig. \ref{pa_cfg}(a) shows the number of CWEs for each code base before (denoted by ``Original'') and after processing with \texttt{PromSec}. Next, Fig. \ref{pa_cfg}(b) presents a histogram of the CWEs in the code bases before and after applying \texttt{PromSec}. It is clearly noticed from both figures that \texttt{PromSec} resolves the majority of CWEs within 20 iterations. In Appendix \ref{appendixA}.3, we trace the evolution in prompt and code and how a CWE is resolved through a toy example.

\par
\begin{figure}[!t]
\centering
\resizebox{0.99\columnwidth}{!}{
\begin{tabular}{cc}
\includegraphics[width=7cm]{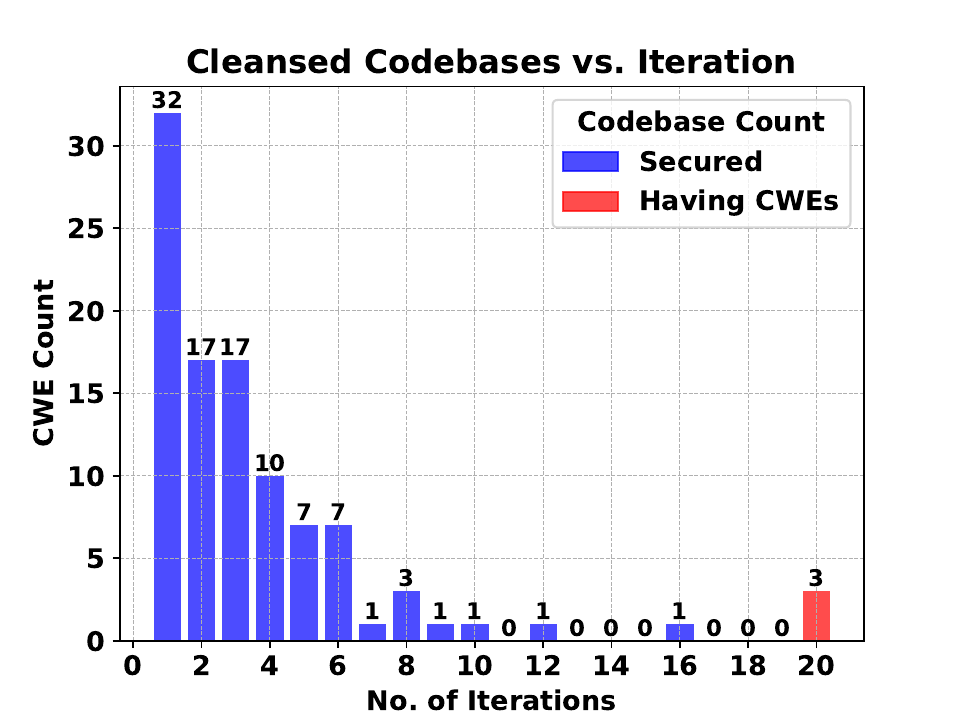}
& 
\includegraphics[width=7cm]{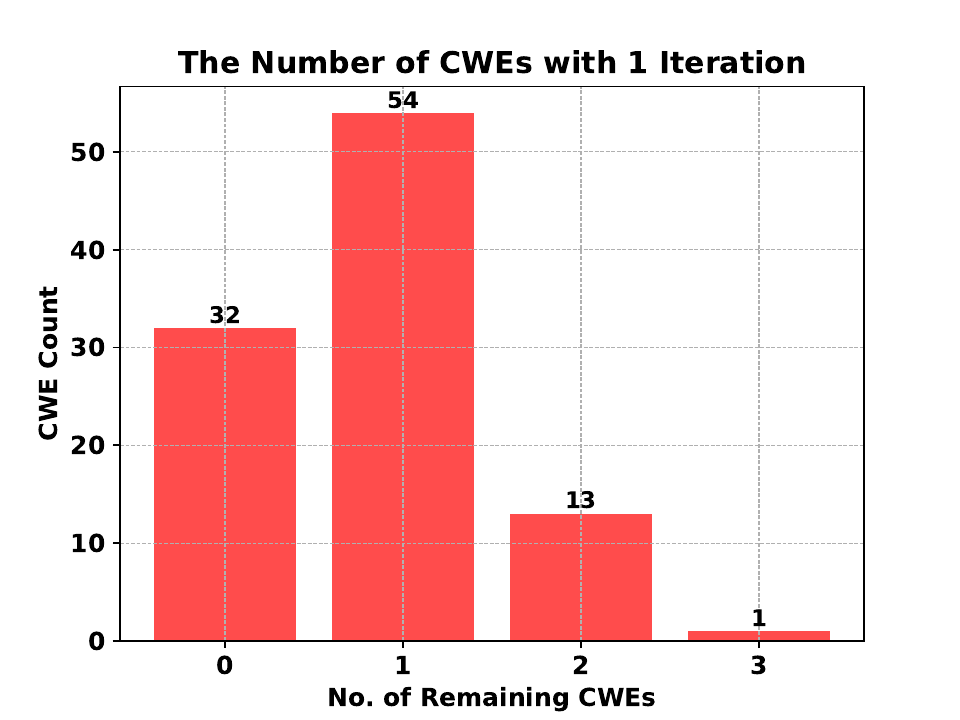}
\\
\Huge{(a)}
&
\Huge{(b)}\\
\end{tabular}}
\vspace{-0.7em}
\caption{(a) The number of cleansed (fully secured) codebases versus iteration (b) the count of remaining CWEs after the first iteration with CFG graphs.}
\label{interations_cfg} 
\end{figure}

 Next, we address \textbf{Q2: Does \texttt{PromSec} need iterative LLM interaction?} For this purpose, we plot a histogram of the number of \textit{secured} code bases versus iteration reporting at the last iteration the number of code bases that still have at least one CWE in Fig. \ref{interations_cfg} (a). We also plot a histogram of the count of code bases that still have remaining CWEs after one \texttt{PromSec} iteration in Fig. \ref{interations_cfg}(b). The two histograms confirm that one iteration is not sufficient to secure the majority of code bases (68\%). On the other hand, around 49\% of the code bases are fully cleaned from CWEs within the first two iterations. 

\begin{figure}[!b]
\centering
\resizebox{0.999\columnwidth}{!}{
\begin{tabular}{c}
\includegraphics[width=15cm]{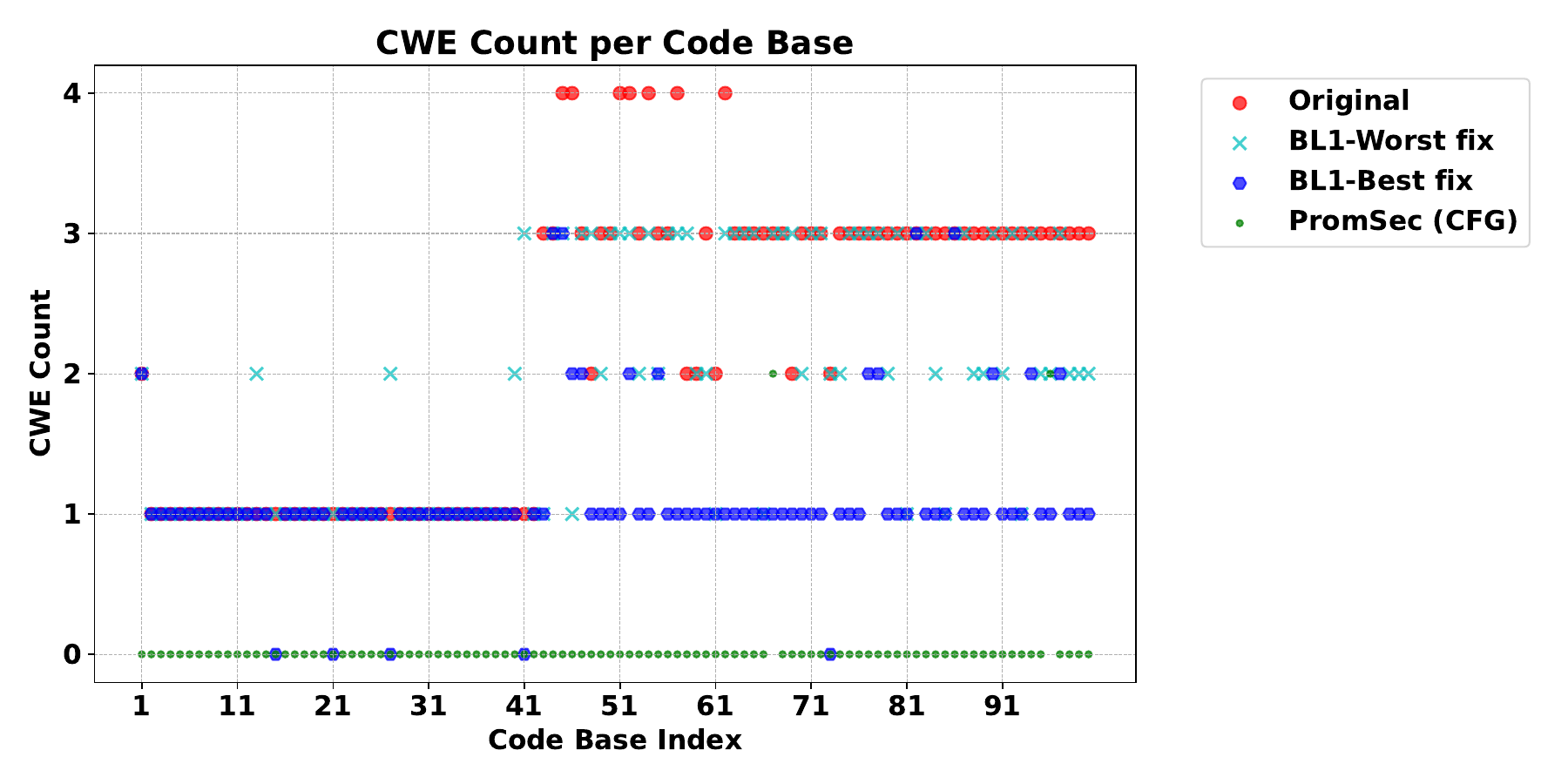}
\\
\Huge{(a)}
\\
\includegraphics[width=15cm]{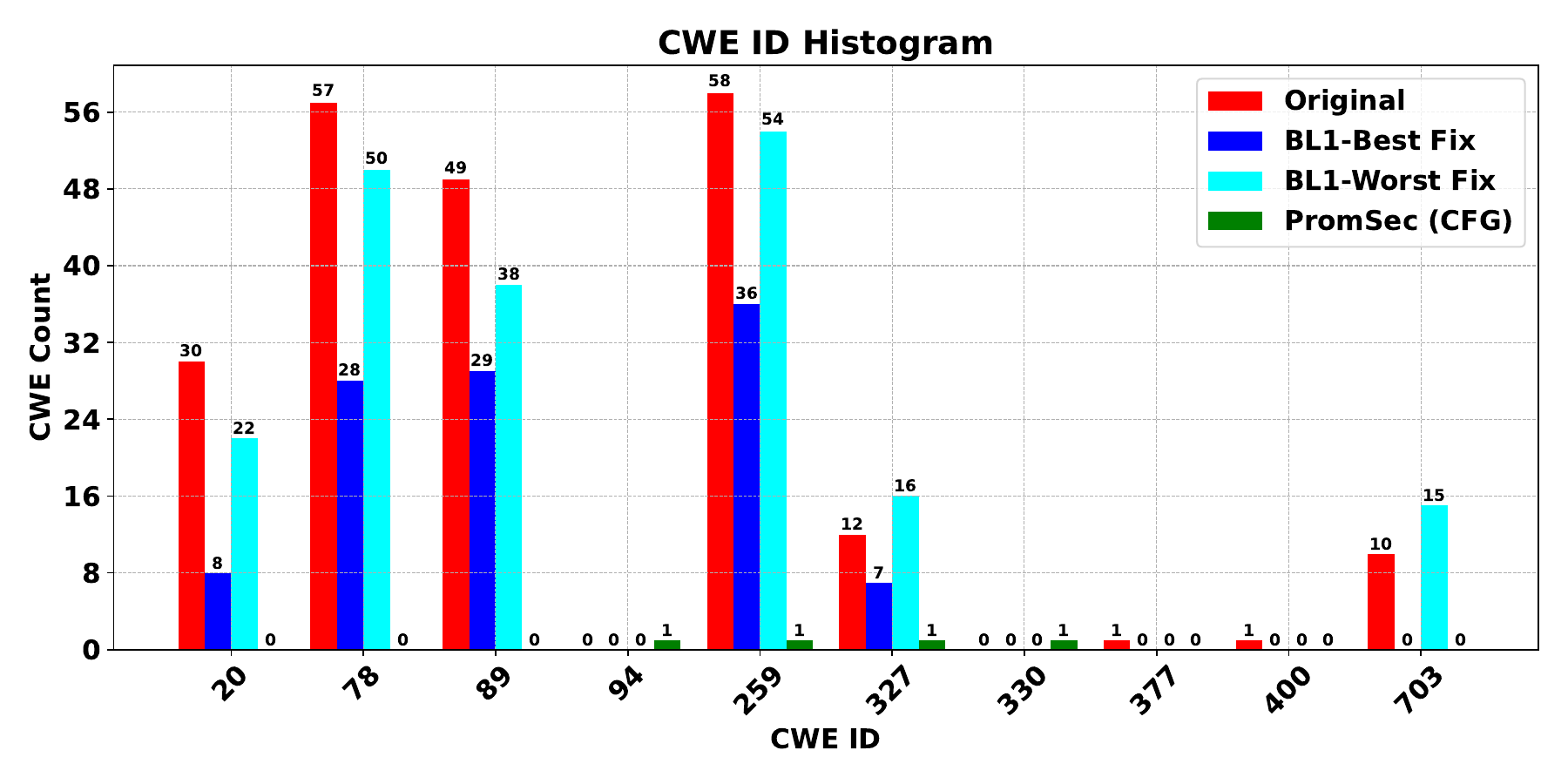}
\\
\Huge{(b)} 
\\
\includegraphics[width=15cm]{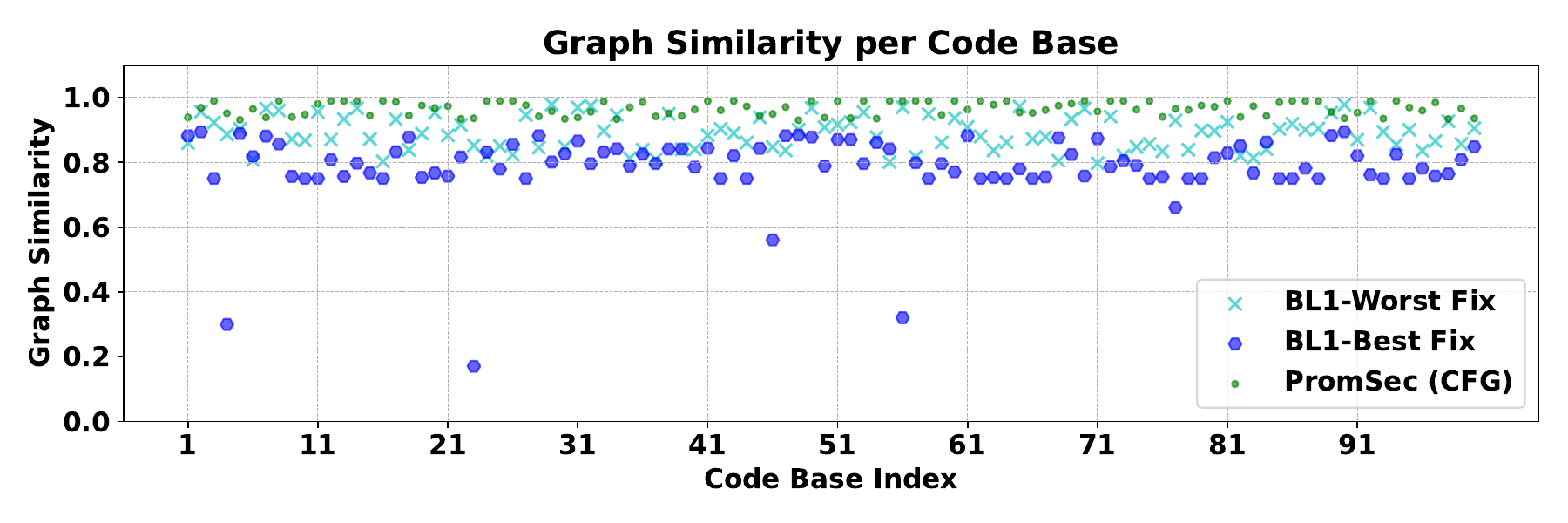}
\\
\Huge{(c)}
\end{tabular}
}
\vspace{-0.7em}
\caption{(a) Comparing \texttt{PromSec} and BL1 in terms of per-code CWE counts, (b) remaining CWE distribution, and (c) code graph similarity.}
\label{SoTA_comp_BL1}
\end{figure}
\par To answer \textbf{Q3: How does \texttt{PromSec} perform relative to the baselines regarding security, functionality preservation, and cost?} we replicate the experiment of Q1 and compare \texttt{PromSec}'s performance with that of the Baseline 1 (BL1). The results are shown in Fig. \ref{SoTA_comp_BL1}. The number of CWEs per code base is shown in \ref{SoTA_comp_BL1}(a), the CWE distribution in \ref{SoTA_comp_BL1}(b), and the comparison of the graph similarity metric (a proxy for functionality preservation) in Fig. \ref{SoTA_comp_BL1}(c).

\par We observe how well BL1 does in the best and worst scenarios in terms of reducing the number of CWEs. In the \textit{best} scenario, we chose the prompt template in BL1 that leads to the fewest CWEs. Conversely, in the \textit{worst} scenario, we chose the one that leads to the maximum CWE count. The results show that \texttt{PromSec} is significantly better than BL1 in resolving CWEs. BL1 does fix some CWEs, but the code it produces is still not completely secure. In terms of preserving code functionality, especially in its best situation, BL1 does not do as well as \texttt{PromSec}. This result shows that \texttt{PromSec} has superior performance both in fixing security bugs (CWEs) as well as upholding the intended code functionality.

\begin{figure}[!thb]
\centering
\resizebox{0.76\columnwidth}{!}{
\includegraphics{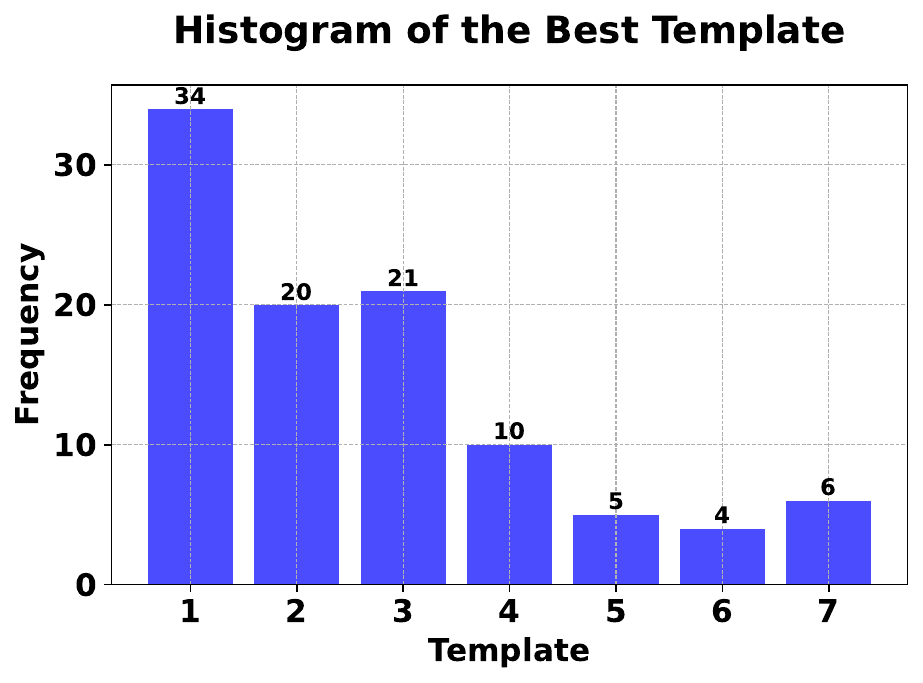}
}
\vspace{-0.7em}
\caption{The best template distribution across code bases for BL1. For 34\% of the codes, no context is required.}
\label{best_template} 
\end{figure}

\begin{figure}[!bt]
\centering
\resizebox{0.99\columnwidth}{!}{
\begin{tabular}{c}
\includegraphics[width=15cm]{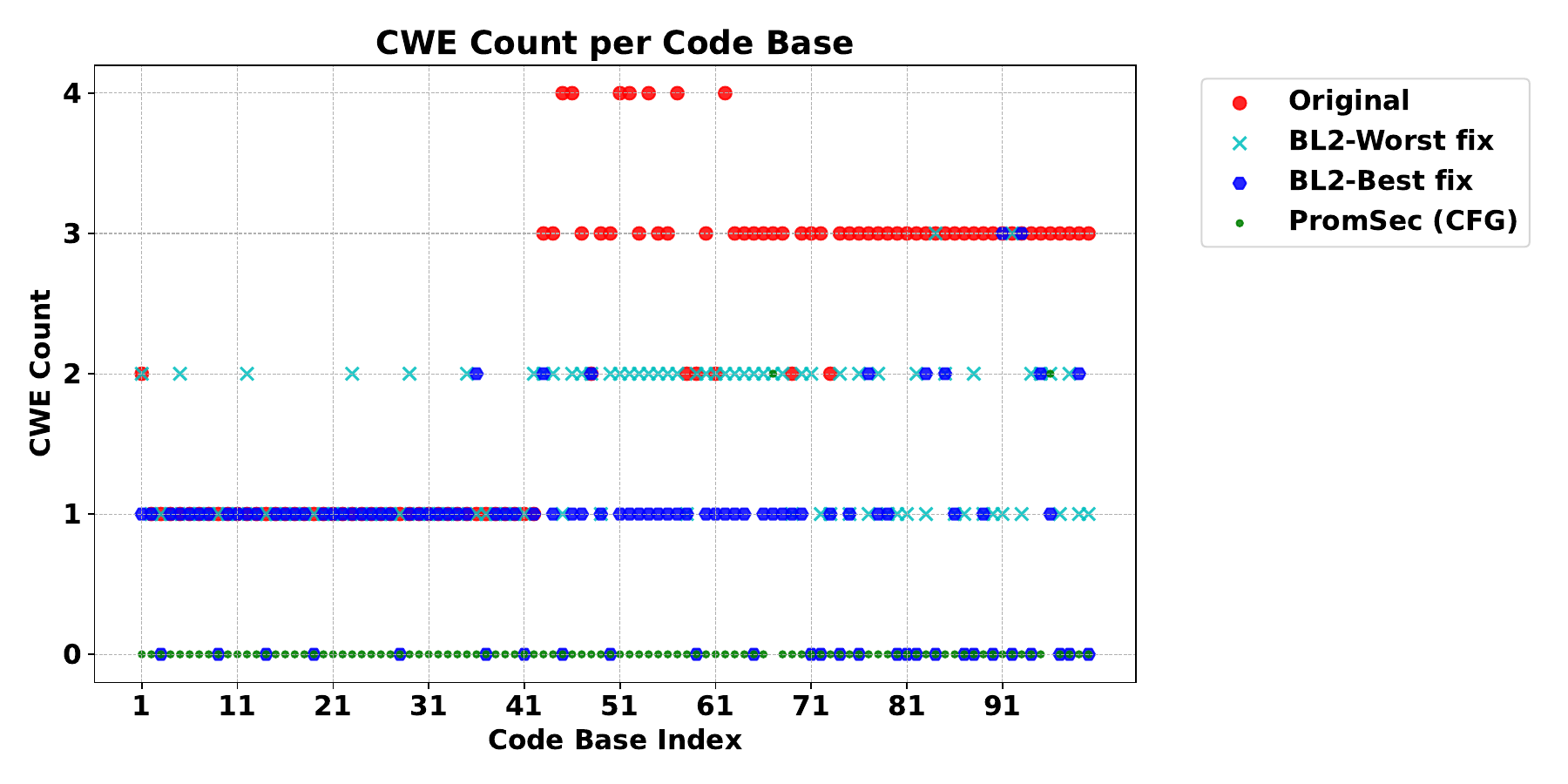}
\\
\Huge{(a)}
\\
\includegraphics[width=15cm]{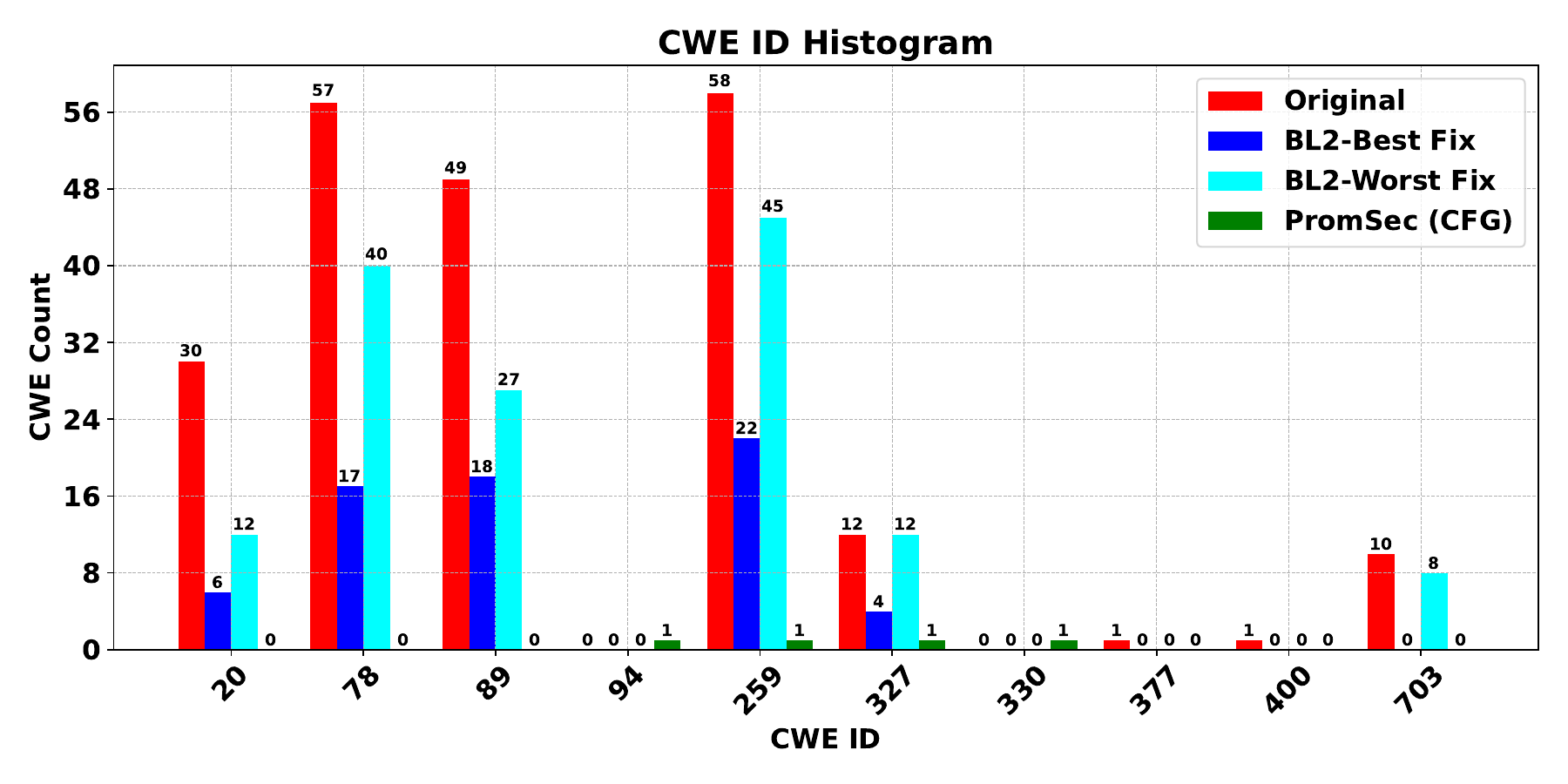}
\\
\Huge{(b)}
\\
\includegraphics[width=15cm]{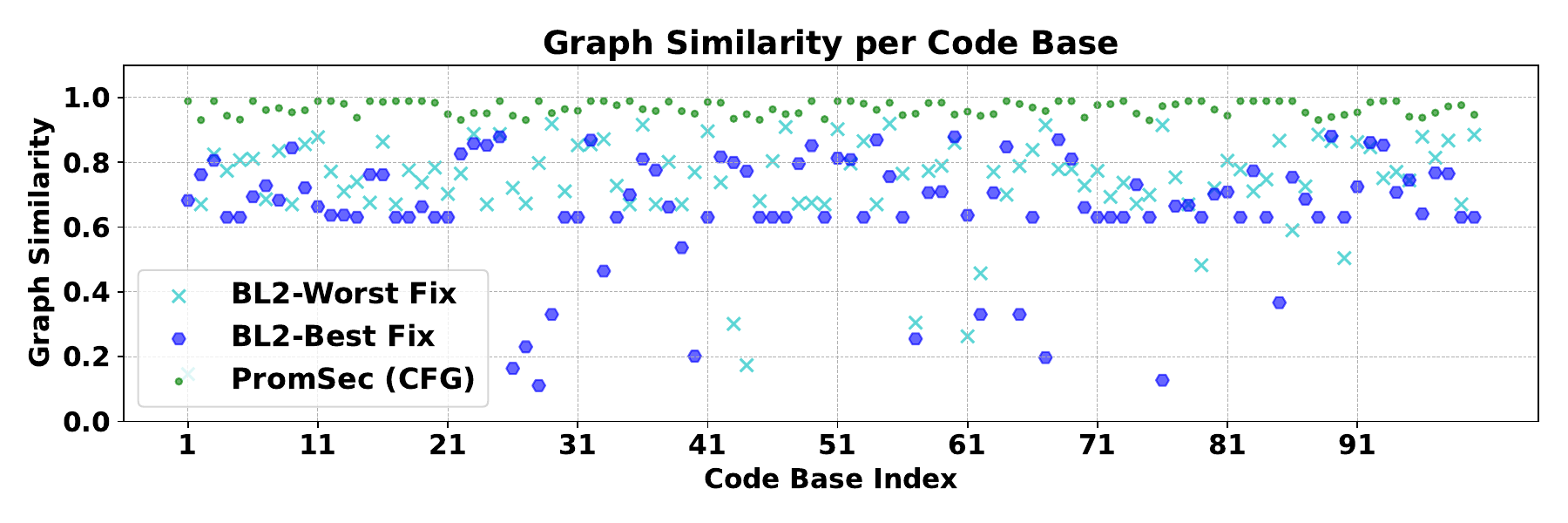}
\\
\Huge{(c)}
\end{tabular}}\\
\vspace{-0.7em}
\caption{(a) Comparing \texttt{PromSec} with BL2 in terms of per-code CWE count, (b) remaining CWE distribution, and (c) code graph similarity.}
\label{SoTA_comp_BL2} 
\end{figure}

\par Pearce et al. \cite{pearce2023examining} (BL1) report that adding more context does not always make code completions more secure. To examine this observation in code generation, we show in Fig. \ref{best_template} how often each prompt template is the most effective for a given test code. Note that we index the templates according to the increase in the context. The histogram in this figure reveals an interesting observation: Template 1, which does not use any security report context, turns out to be the best choice 33\% of the time. This result suggests that while including some context is beneficial, adding too much does not always help. In fact, we notice a decrease in preference for templates with more context. This observation is consistent with Pearce et al.’s observation in code completion. 

\par Next, we compare the effectiveness of \texttt{PromSec} against the other baseline, BL2, and present the results in Fig. \ref{SoTA_comp_BL2}. This figure reveals that BL2 outperforms BL1 in terms of security enhancement, as evidenced by a notably lower count of remaining CWEs. However, this improvement in security enforcement comes at the expense of a reduction in code functionality preservation as shown in Fig. \ref{SoTA_comp_BL2}(c), where BL2 results in a greater degeneration in code functionality.

\par \textbf{Fuzzing Tests.} In this experiment, we address \textbf{Q4: How effective is \texttt{PromSec} in preserving the intended code functionality?} In software engineering, fuzzing testing (FT) \cite{li2018fuzzing} is a procedure typically used to compare the functionality of two source codes. It involves giving the same inputs to both codes and then measuring differences in their outputs. However, FT usually requires a lot of manual work and can be time-consuming, especially for large numbers of code bases. Also, FT only works well when the code bases have clear and measurable inputs and outputs. Because of these limitations, we use FT as a secondary method to check how well \texttt{PromSec} preserves the original semantics and function of code while improving its security. We compare \texttt{PromSec} with the BL1 and show the results in Table \ref{fuzzing_test}.

\par For conducting FT, we use 20 specially designed code bases that can be tested with fuzzing. Each code base has a main function with specific input and output parameters and calls several internal functions. This setup allows us to quantify any changes in the outputs caused by \texttt{PromSec}. We repeat these tests for 1,000 trials. In each trial, we sample random inputs and calculate the differences in outputs. Then, we calculate the average value of the output differences. An FT is passed if the output changes on average, are smaller than a small threshold value (we set it to 0.01 to account for minor numeric variations
from floating-point calculations, ensuring functional equivalence). Despite this
threshold, differences are technically zero in the experiments.

\begin{table}[t]
\small
\centering
\caption{FT comparison of \texttt{PromSec}, BL1, and BL2. Code bases are fully secured by each method, and the number of FT passes is shown.}
\vspace{-0.7em}
\begin{tabular}{|c|c|c|c|c|}
\hline
\multirow{2}{*}{Initial $k$} & \multirow{2}{*}{ No. of Code bases} & \multicolumn{3}{c|}{No. of Passes} \\ \cline{3-5} 
            &                & PromSec & BL1 & BL2 \\ \hline
78          & 10             & 10           & 6        & 3        \\ \hline
89          & 10             & 10           & 5        & 2        \\ \hline
Overall     & 20             & 20           & 11       & 5       \\ \hline
\end{tabular}
\label{fuzzing_test}
\end{table}

\par In Table \ref{fuzzing_test} we process a total of 20 code bases that exhibit CWEs 78 and/or 89 with \texttt{PromSec}, BL1, and BL2. Security-wise all these methods fully secure the code bases in the experiment. The table shows that \texttt{PromSec} fully passes the test for all the samples considered. On the other hand, BL1 exhibits some degree of functionality preservation but lower than that achieved by \texttt{PromSec}, whereas BL2 severely fails. This is consistent with the finding on BL2's functionality preservation in the previous experiment.

\begin{table}[t]
\centering
\caption{Empirical cost of BL1, \texttt{PromSec}(1), BL2, and \texttt{PromSec}(20).}
\vspace{-0.7em}
\resizebox{0.99\columnwidth}{!}{
\begin{tabular}{|cl|c|c|c|c|}
\hline
\multicolumn{2}{|c|}{Cost Aspect} & BL1 & \texttt{PromSec}(1) & BL2 & \texttt{PromSec}(20) \\ \hline
\multicolumn{1}{|c|}{Time Cost (seconds)} & Overall Time & 35.62 & 9.95 & 649.16 & 70.92 \\ \hline
\multicolumn{1}{|c|}{\multirow{4}{*}{LLM Query Cost}} & Number of LLM Queries & 7.00 & 2.00 & 127.48 & 14.36 \\ \cline{2-6} 
\multicolumn{1}{|c|}{} & Input Token Count & 1153.10 & 359.20 & 22755.80 & 2359.94 \\ \cline{2-6} 
\multicolumn{1}{|c|}{} & Output Token Count & 574.90 & 256.00 & 10348.20 & 1904.46 \\ \cline{2-6} 
\multicolumn{1}{|c|}{} & LLM Model Query Time & 33.94 & 9.50 & 604.92 & 66.70 \\ \hline
\multicolumn{1}{|c|}{\multirow{2}{*}{Security Analysis Cost}} & Number of Security Analyses & 7.00 & 1.00 & 127.48 & 6.92 \\ \cline{2-6} 
\multicolumn{1}{|c|}{} & Security Analysis Time & 1.652 & 0.237 & 29.738 & 1.756 \\ \hline
\end{tabular}}
\label{empir_cost_table}
\end{table}

\par \textbf{Empirical Cost Analysis} The third aspect of comparing \texttt{PromSec} to the baselines is the operational cost. We compare the costs of BL1 along with one iteration of \texttt{PromSec} denoted by (\texttt{PromSec}(1)), 
and BL2 with a 20-iteration \texttt{PromSec} (denoted by \texttt{PromSec}(10)). The results are shown in Table \ref{empir_cost_table}. As seen in the table, \texttt{PromSec}(1) demonstrates improvements over BL1. The overall time cost is notably lower with \texttt{PromSec}(1), with a mean of 9.95 seconds, in contrast to 70.92 seconds. The LLM query cost category shows a reduction in the number of queries, input and output token counts, and LLM model query time, with \texttt{PromSec}(1) requiring 2 queries versus 7 in BL1, and token counts of 359.2 and 256 compared to 1,153.1 and 574.9, respectively. The query time is also reduced to 9.5 seconds from 33.94 seconds. In the security analysis cost, \texttt{PromSec}(1) requires fewer security analysis queries (1 vs. 7) and less time (0.237 seconds vs. 1.652 seconds). These results indicate that \texttt{PromSec}(1) is more efficient in terms of both time and resource management compared to the existing BL1 baseline. It is noted that this is still the case even though \texttt{PromSec} needs on average 3 iterations, since even with multiplying the cost of \texttt{PromSec}(1) by 3, it is still superior to that of BL1. A more obvious advantage is observed for \texttt{PromSec}(20) over BL2, where \texttt{PromSec}(20) has around an order of magnitude reduction compared with BL2.

\subsection{Different Code Graph Types}
\par
\begin{figure}[!ht]
\centering
\resizebox{0.99\columnwidth}{!}{
\begin{tabular}{c}
\includegraphics[width=15cm]{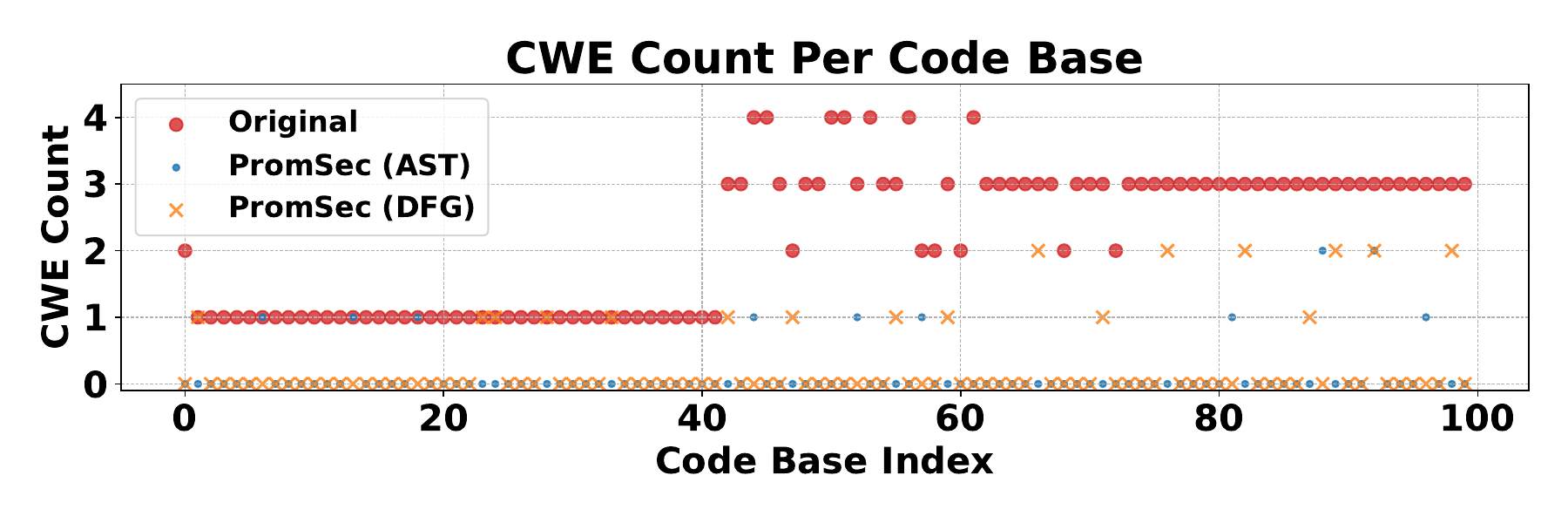}
\\
\Huge{(a)}
\\
\includegraphics[width=15cm]{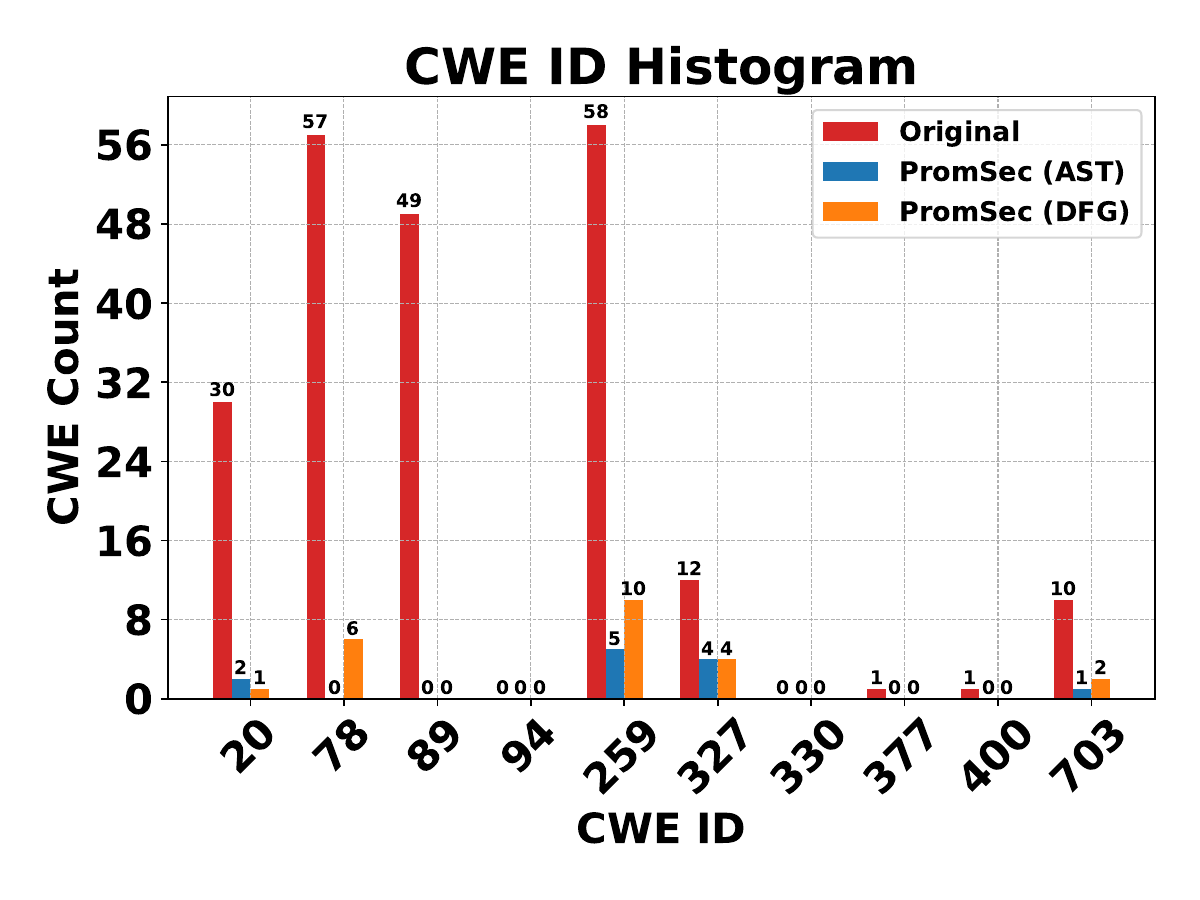}
\\
\Huge{(b)}
\end{tabular}}
\vspace{-0.7em}
\caption{Performance evaluation with AST- and DFG-type graphs: (a) per code base CWE count before and after applying \texttt{PromSec} and (b) CWE counts before and after applying \texttt{PromSec}.}
\label{pa_ast} 
\end{figure}

 Here, we answer \textbf{Q5: What is the impact of the code graph type on performance?} We evaluate the performance of \texttt{PromSec} when it utilizes either AST or DFG graph types instead of CFG. Fig. \ref{pa_ast} shows the performance analysis based on AST and DFG graphs. First, consider the AST performance, we note that this performance is acceptable but inferior when compared to the scenario involving CFG graphs. Part of the reason for this is that although ASTs encapsulate the code's syntax, they lack the depth needed to encompass control flow and operational semantics, crucial for a comprehensive security analysis. On the other hand, the performance with DFG-type code graphs is inferior to that of AST-type graphs. Also, both are inferior to CFG graphs. DFGs focus on data flows and dependencies but do not capture the logic and behavior of codes as done by CFGs.



\subsection{\texttt{PromSec}'s Transferability to Unseen CWEs}
\par As shown in our experiments, the ability of the trained gGAN model to fix CWEs is established. Now, we address \textbf{Q6: Can \texttt{PromSec} transfer to unforeseen CWEs in its training process?} For this purpose, we conduct the following experiment. We exclude a given test CWE from the training set of the gGAN model and refer to it as a \textit{masked} CWE. Then, we pick test examples that exhibit this masked CWE. In other words, the model does not see any training examples exhibiting the masked CWE.

\par
\begin{figure}[!t]
\centering
\resizebox{0.99\columnwidth}{!}{
\includegraphics{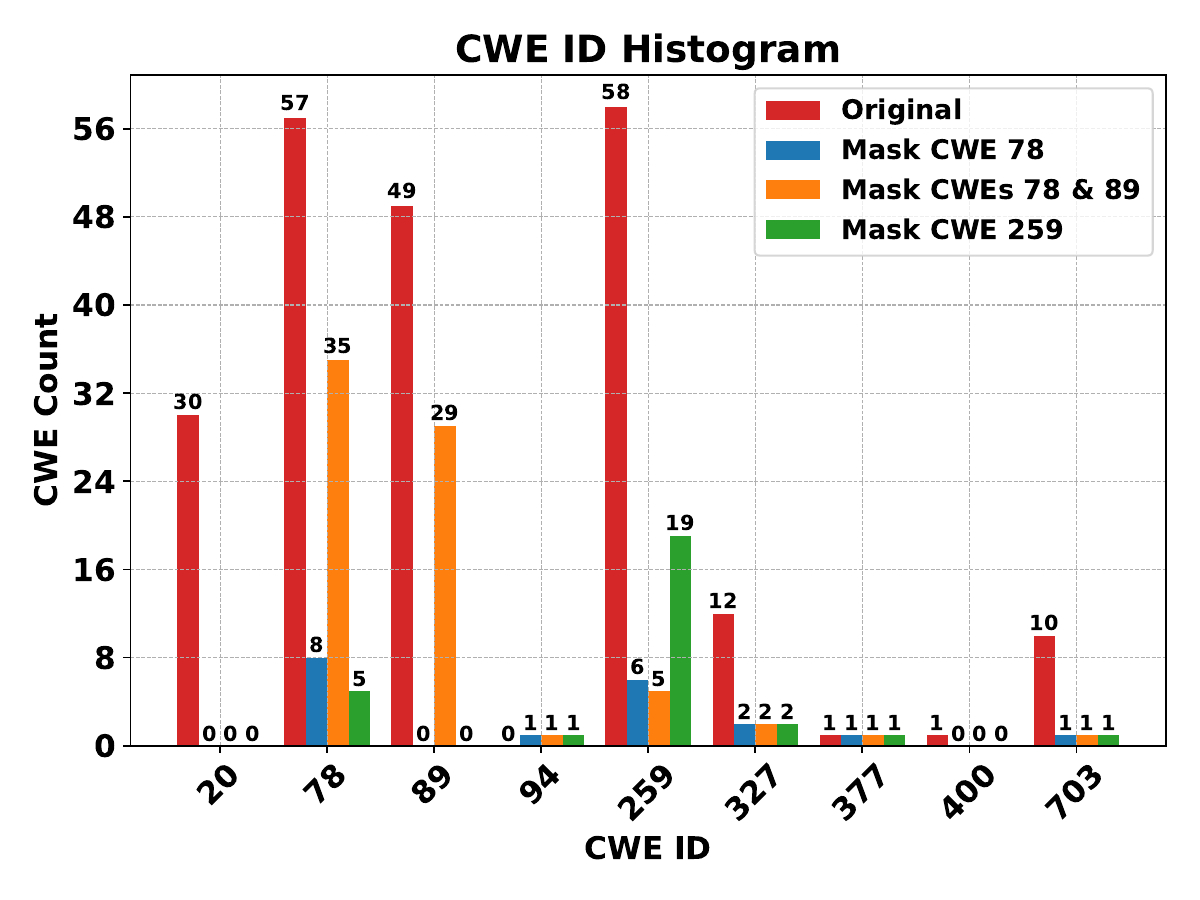}
}
\vspace{-0.7em}
\caption{Cross-CWE transferability: The histogram CWEs before and after applying \texttt{PromSec} with masking: CWE 78, both CWE 78 and CWE 89, and CWE 259.}
\label{masking_cwes} 
\end{figure}

 Fig. \ref{masking_cwes} depicts the effects of masking specific CWEs, namely, CWE 259, CWE 89, and CWE 78, on \texttt{PromSec}'s performance. Several observations can be drawn from this figure. First, masking a given CWE results in a slight decrease in \texttt{PromSec}'s effectiveness against it and a minor reduction in handling some other related CWEs. This observation validates \texttt{PromSec}'s ability to transfer its security-enhancing capabilities to CWEs not directly included in its training. We attribute this transferability to the generalization capability of the gGAN model. Furthermore, it is intuitive to assume that resolving a specific CWE may have a beneficial impact on mitigating related CWEs, especially if they originate from similar underlying issues. For example, prompt optimizations directed to prevent CWE 78 (input injection) might implicitly also mitigate CWE 89 (SQL injection) risks. This result is because both vulnerabilities typically stem from inadequate control of user input, whether it be in operating system commands (in CWE 78) or database queries (in CWE 89). This analysis hints at the possibility of mitigating new security vulnerabilities not explicitly included in the training process of the gGAN.

\subsection{\texttt{PromSec}'s Transferability across LLMs}
\par
\begin{figure}[!tb]
\centering
\resizebox{0.99\columnwidth}{!}{
\begin{tabular}{c}
\includegraphics[width=15cm]{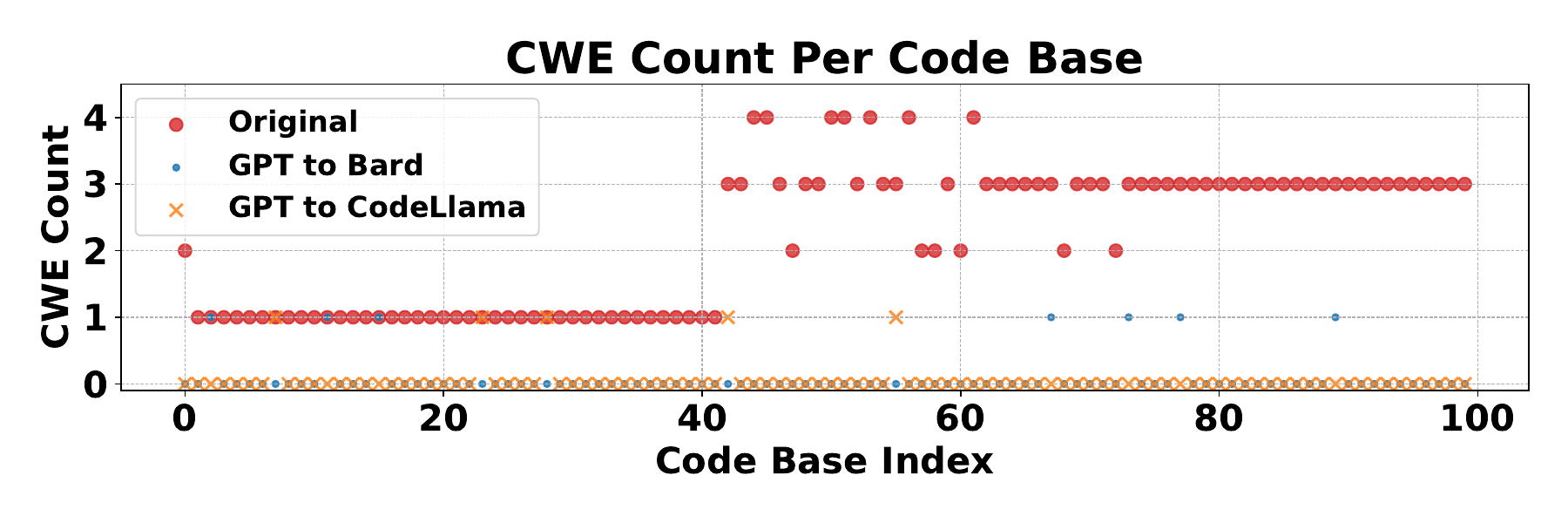}
\\
\Huge{(a)}
\\
\includegraphics[width=15cm]{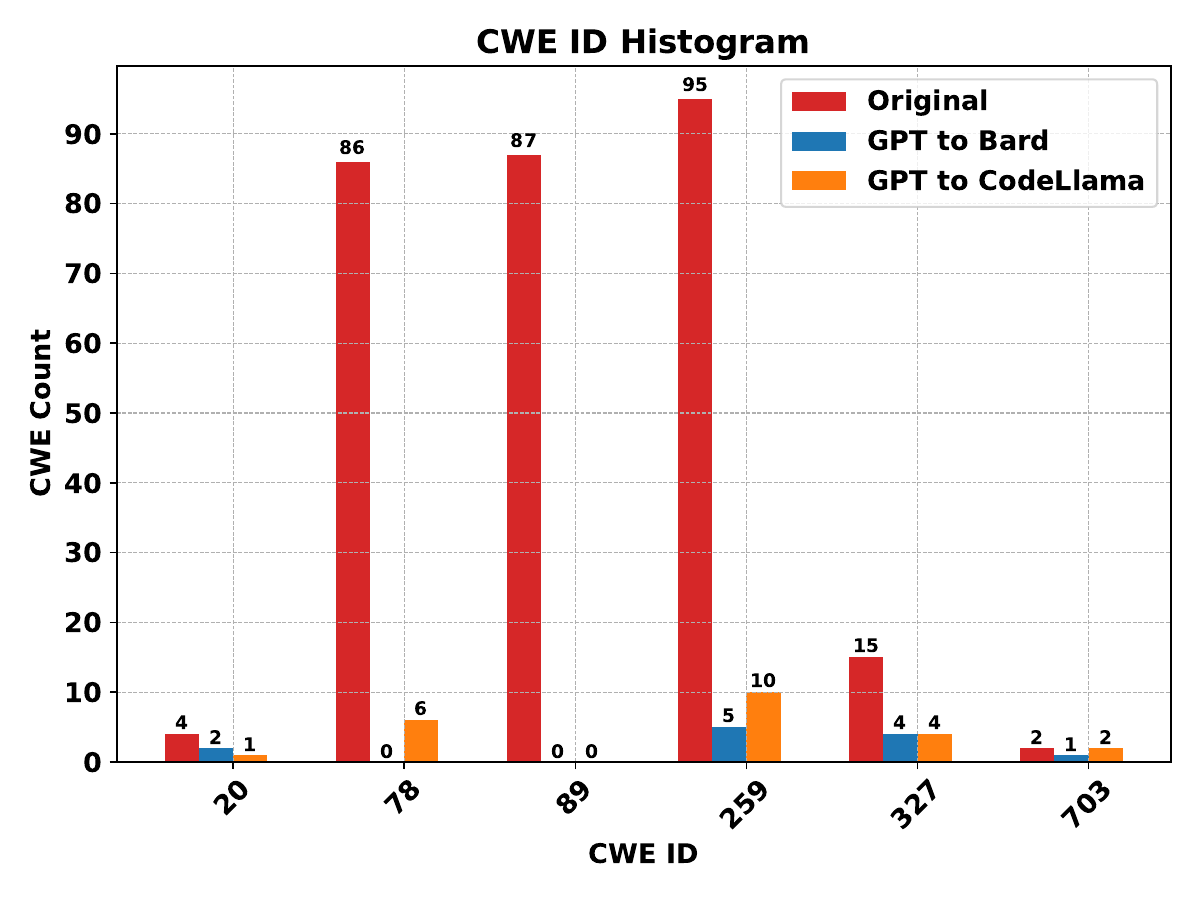}
\\
\Huge{(b)}
\end{tabular}}\\
\vspace{-0.7em}
\caption{Cross-LLM Transferability: Prompts optimized with GPT-3.5 Turbo are used to generate code with Bard (GPT-to-BARD) and Code Llama (GPT-to-CodeLlama). (a) CWE counts per code base, and (b) CWE counts before and after applying \texttt{PromSec}.}
\label{transfer_bard} 
\end{figure}

 Considering the pipeline in Fig. \ref{overall_system}, we address \textbf{Q7: Are prompt fixes obtained with a given LLM exclusively valid for that LLM or can they be transferable to other LLMs?} For this purpose, we carry out the following experiment. We assume GPT-3.5 Turbo as the base LLM used for optimizing the prompts while we feed these optimized prompts to another LLM to generate the code. We consider Google's Bard and Meta's CodeLlama-13B-Instruct as these LLMs and show the corresponding results in Fig. \ref{transfer_bard}. In both figures, prompts optimized with GPT-3.5 Turbo work well with both Bard and CodeLlama-13B-Instruct. The performance exhibits negligibly small remaining CWEs. This result validates the ability of \texttt{PromSec}'s prompt optimizations to transfer across LLMs.

\subsection{\texttt{PromSec}'s Transferability across Programming Languages}

\par
\begin{figure}[!hbt]
\centering
\resizebox{0.99\columnwidth}{!}{
\begin{tabular}{c}
\includegraphics[width=15cm]{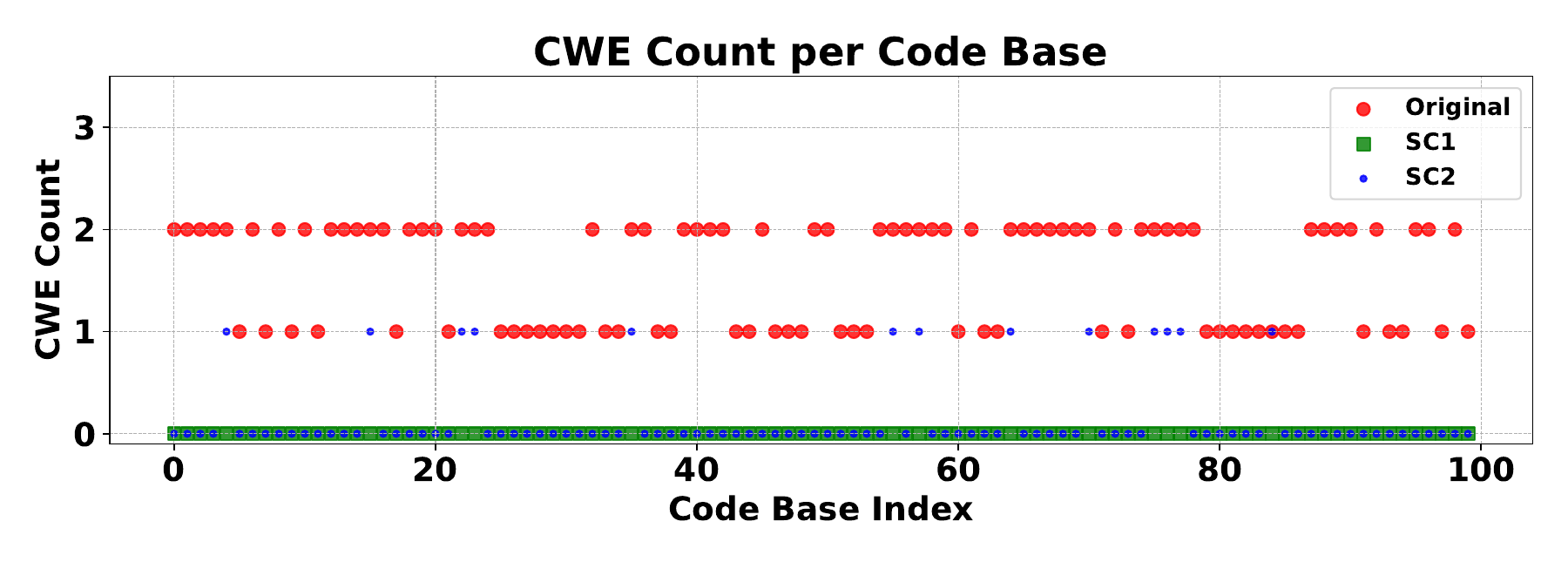}
\\
\Huge{(a)}
\\
\includegraphics[width=15cm]{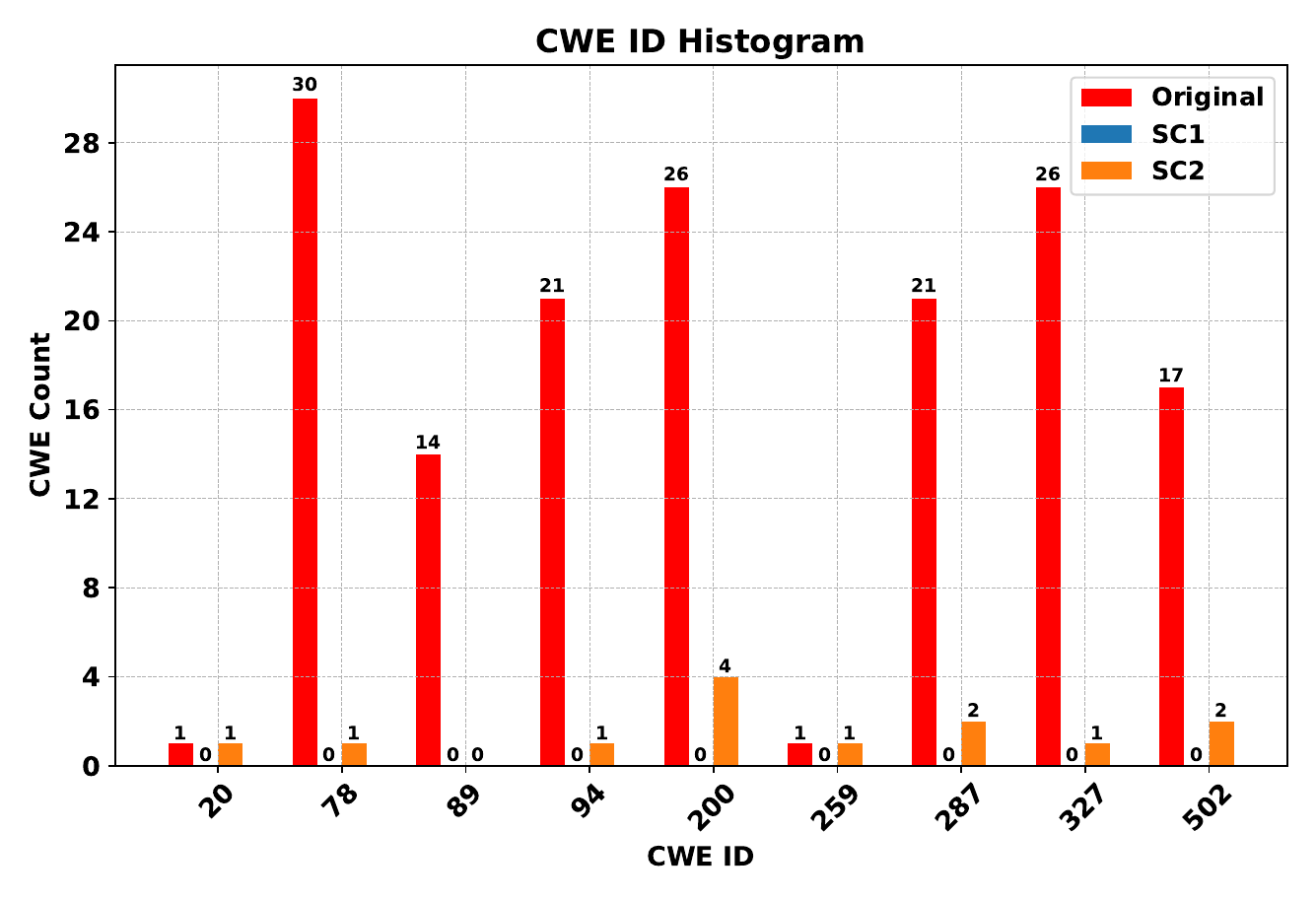}
\\
\Huge{(b)}
\\
\includegraphics[width=15cm]{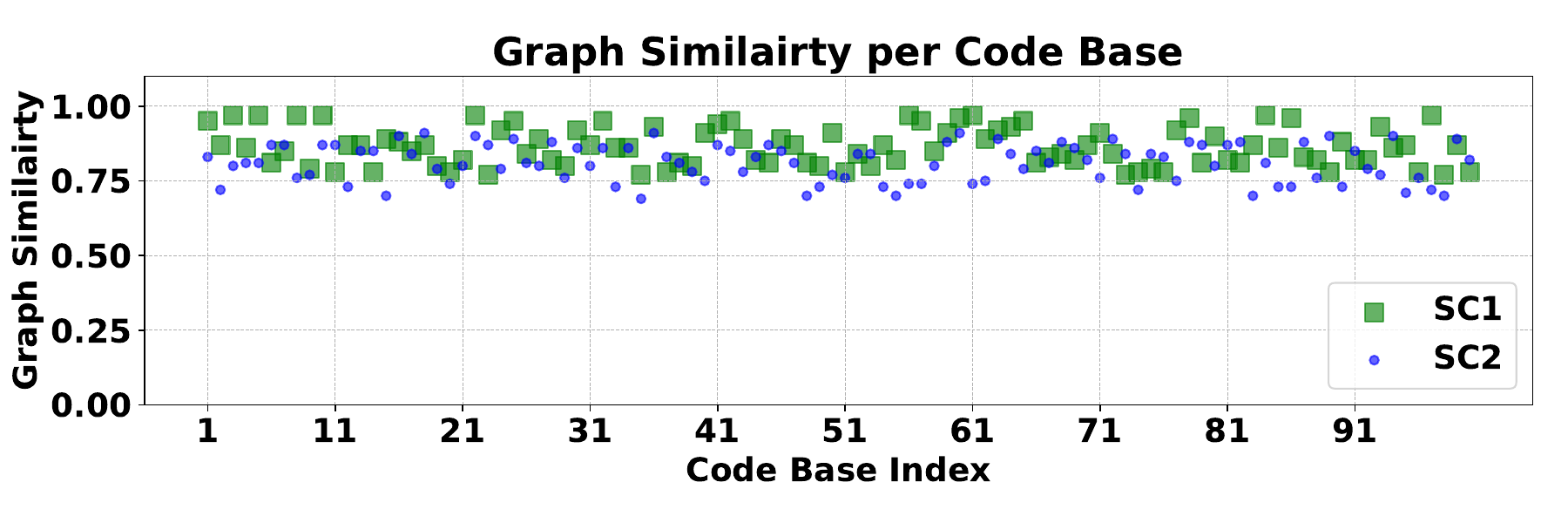}
\\
\Huge{(c)}
\end{tabular}}\\
\vspace{-0.7em}
\caption{For SC1 (training and testing with Java), and SC2 (training with Python and testing with Java): (a) the CWE count before and after applying \texttt{PromSec}, (b) the CWE ID histograms before and after applying \texttt{PromSec}, and (c) graph similarity.}
\label{java} 
\end{figure}

 In this subsection we answer \textbf{Q8: How successfully can \texttt{PromSec}'s optimized prompts generalize across different programming languages?} For this purpose, we collect a dataset of Java prompts from \cite{hao2022aixbench,yu2023codereval}. Next, we consider these prompts as inputs to \texttt{PromSec}’s pipeline illustrated in Fig. \ref{overall_system}. We intentionally employ these prompts to examine the operation of this pipeline starting with prompts, rather than source code as done in the tests considered so far. Next, we train a gGAN model over 200 code bases generated from these prompts. After that, we use the remaining 100 prompts to test the operation of \texttt{PromSec} (we denote this scenario by SC1). To examine the generalization of \texttt{PromSec} across programming languages, we test this prompt dataset with the previous gGAN trained on Python code (we denote this scenario by SC2). Fig. \ref{java} shows a histogram of the CWEs in the test set before and after applying \texttt{PromSec} for these two scenarios in (a) and (b), respectively. Fig\ref{java}(a) clearly shows that \texttt{PromSec} perfectly resolves the CWEs, similar to the case with Python code experiments. Note that we use the same hyperparameters and number of iterations. Next, Fig\ref{java}(b) shows almost the same performance is attainable with \texttt{PromSec} despite being trained using codes from another language. The result in Fig\ref{java}(b) showcases the transferability of \texttt{PromSec} across source code of different programming languages. 

\subsection{An Ablation Study}
\par
\begin{figure}[!ht]
\centering
\resizebox{0.78\columnwidth}{!}{
\includegraphics[width=25cm]{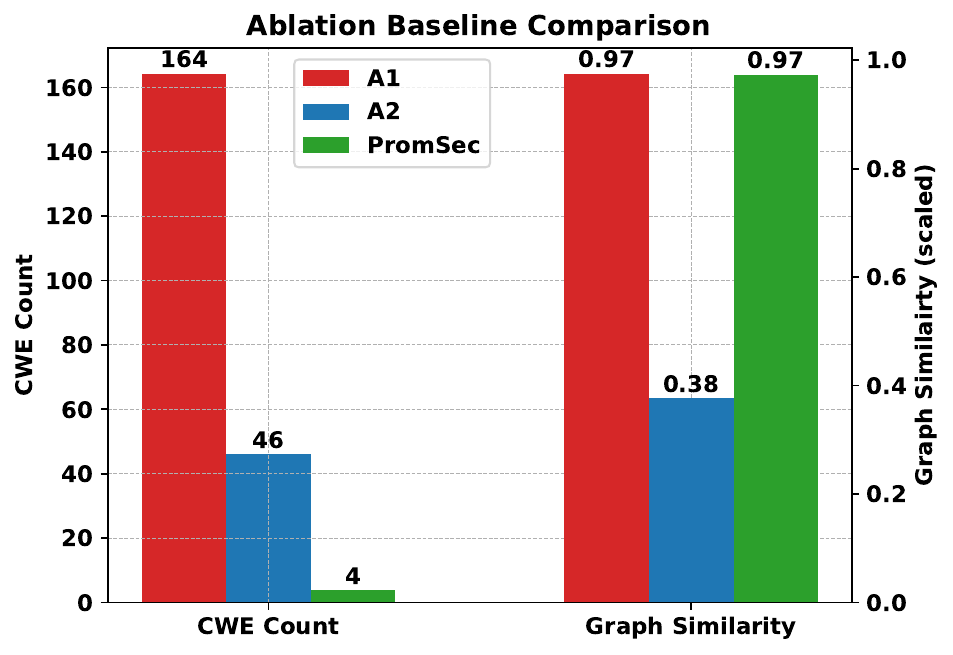}
}
\vspace{-0.7em}
\caption{An ablation study comparing CWE counts and code graph similarity across \texttt{PromSec}, scenario A1 (without gGAN), and scenario A2 (without LLM).}
\label{combined_ablation} 
\end{figure}

 To examine the individual roles of the gGAN and the LLM components in the \texttt{PromSec} pipeline, we conduct ablation studies comparing \texttt{PromSec} to two scenarios. In the first scenario (A1), \texttt{PromSec} operates without the gGAN, iterating solely with the LLM. In the second scenario (A2), we exclude the LLM, applying only the gGAN in the pipeline. The outcomes are presented in Fig. \ref{combined_ablation}. As shown, excluding the gGAN significantly impairs the pipeline’s ability to reduce CWE counts while maintaining code functionality, indicated by high code graph similarity scores. Conversely, omitting the LLM does not notably affect CWE reduction but significantly undermines code functionality, demonstrated by reduced code graph similarity.

\par The drop in graph similarity without the LLM is due to the gGAN's lack of contextual understanding. LLMs capture and preserve the code’s semantics and functionality, continuously emphasizing its intent. Without LLMs, the gGAN alters code graphs recursively, losing original functionality. LLMs ensure prompts reflect and emphasize the code’s functionality and structure, guiding the gGAN to produce precise outputs. These results highlight the interplay between the two components in achieving \texttt{PromSec}'s goal: enhancing code security without sacrificing functionality.

\section{Related Work}
\label{Section6}

\par This section briefly revisits prompt optimization for LLMs, especially for secure code generation, and GNNs for code analysis.

\par \textbf{Prompt Optimization for General LLM Tasks.} Prompt engineering and optimization are crucial in maximizing the potential of LLMs. Works in this area vary by the assumed degree of LLM access and prompt treatment methods. Broadly, they can be divided into two main categories \cite{liu2023pre}. 
The first category considers soft prompts. These works consider a continuous prompt space represented as embeddings for the model \cite{qin2021learning,liu2023gpt,shin2020autoprompt}. This approach involves employing differentiable prompts for these soft embeddings. However, these studies presume familiarity with the latent embeddings of the LLM, leading to limitations in scalability. The second category involves hard prompts and adopts a discrete prompt treatment. This approach starts with an initial prompt and focuses on prompt engineering or optimization by generating a candidate set of feasible prompts and selecting the most suitable one. Examples along this line include the use of Monte-Carlo sampling by Zho et al. \cite{zhou2022large}, gradient-based and beam search by Pryzant et al. \cite{pryzant2023automatic}, and genetic algorithm by Xu et al. \cite{xu2022gps}. However, these methods, while accomplishing some success, remain computationally intensive and lack assured convergence guarantees. Li et al. \cite{li2023guiding} offer another instance where reinforcement learning agents dynamically enhance provided prompts within textual LLM tasks, like translation and text summarization. While these endeavors advance prompt engineering, they underscore the persistent hurdle of crafting universally effective prompts. However, these efforts do not address prompt optimization for utilizing LLMs for source code generation. Furthermore, none of these studies delve into optimizing prompts for bolstering security measures.

\par \textbf{Prompt Engineering and Optimization for Secure Code Generation.}
Recent research exhibits efforts for securing source code generation with LLMs. Along this line, Pearce et al. \cite{pearce2023examining} use prompt engineering to guide LLMs with manually created prompt templates enriched with context from static security tools. However, this approach struggles with preserving the original functionality of the code and is constrained by the limitations of static security tools \cite{charalambous2023new,gadelha2019smt}. Besides, incorporating more context from security reports does not necessarily improve performance. Charalambous et al. \cite{charalambous2023new} use bounded model checking (BMC) to direct LLMs in generating memory-safe code. While successfully repairing buffer overflow and pointer dereference issues, this method is restricted to fixing memory safety issues detectable by BMCs. Other studies \cite{islam2024code,islam2024llm} use reinforcement learning to enhance the security of code generated by LLMs, focusing on source code instead of prompts. Also, \cite{he2023large} employs trainable prefixes for influencing LLMs in either defense (security hardening) or offense (vulnerability creation). The optimization of a prefix requires access to the model’s parameters and hidden states. It thus restricts the applicability of this method to open-source models, excluding proprietary models like GPT, Bard, and Llama. 
Collectively, these studies underscore the need for more versatile and reliable methods in ensuring the security and functionality desiderata of code generated by LLMs.

\par \textbf{GNNs for Source Code Analysis.} Software engineering has significantly benefited from ML and deep learning advancements, with ASTs commonly used as input data \cite{white2016deep}. GNNs have recently gained attention for semantic code analysis, marking a shift from traditional syntax-based methods. GNNs are effective in representing complex code structures and semantics, aiding in several tasks. In practical applications, GNNs have shown promise in software bug detection and correction. HOPPITY \cite{dinella2020hoppity} uses GNNs for bug detection and fixing through graph transformations. Devgin \cite{zhou2019devign} employs a heterogeneous GNN for vulnerability detection in source code by constructing a heterogeneous code graph. Code clone detection is another area where GNNs excel. They can uncover semantic similarities in clones that appear syntactically different. DeepSim \cite{zhao2018deepsim} uses CFGs for clone detection, while \cite{wang2020detecting,liu2023learning} combine various code graphs like ASTs, CFGs, and DFGs to improve clone detection accuracy. Additionally, \cite{fang2020functional} blends syntactic and semantic information for clone detection. This research body highlights the crucial role of GNNs and graph embeddings in source code analysis.

\section{Discussion}
\label{Section7}
\par \textbf{Impact of the Study.} This research represents a significant step toward the widespread and practical utilization of LLMs in software development at a real-world and large-scale level. The effectiveness of \texttt{PromSec} across various code bases and its potential to generalize to unforeseen vulnerabilities and diverse LLMs highlight its ability to instill confidence in LLMs for secure and dependable code generation. This effectiveness, in turn, paves the way for their expanded application in real-world software development contexts.

\par \textbf{Limitations.} As also acknowledged in other works like \cite{pearce2023examining,he2023large}, a limitation to the validity of this work's findings is the limited coverage of static security tools. Also, despite the adoption of code graph similarity as a proxy quantification for functional similarity \cite{zhao2018deepsim,fang2020functional,wang2020detecting,liu2023learning}, techniques like FT \cite{li2018fuzzing} are more accurate in doing so. Nonetheless, their applicability is challenged by requirements of measurable code inputs and outputs and entails manual preparation of the tests. This limited our ability to apply FT at a large scale. Another issue could be raised about the coverage of the CWEs considered based on the ones exhibited in the training datasets we use. However, our experiment shows strong generalizability and transferability of \texttt{PromSec} which enables it to cover a wider range of unseen CWEs. Finally, in scenarios where \texttt{PromSec} is applied to excessively large code bases with lengths exceeding the LLMs’ context size, it may face a scalability issue. However, it is uncommon for code bases to reach such extensive lengths. A promising solution is dividing large codes into manageable units (based on modular or structural criteria), and then focusing on divisions with vulnerabilities. This approach requires further research on optimizing the division. Also, advancements in LLM’s prompt window widths are expected to mitigate this concern progressively.

\par \textbf{Future Work.} Future research can explore various paths to enhance the applicability and effectiveness of \texttt{PromSec}. A promising extension is to further improve the prompts aiming at reducing the number of \texttt{PromSec} iterations. This can be done by aggregating prompt fixes either from multiple LLMs and/or from multiple gGANs utilizing different types of code graphs. Another extension may involve exploiting the corpus of prompt tuples (the initial prompt, intermediate prompts, and the final prompt) generated by \texttt{PromSec} to train an end-to-end ML-based prompt optimization approach, potentially using a smaller-scale LLM for this purpose. Even though we opted to operate \texttt{PromSec} autonomously, an interesting point is to consider a human-in-the-loop which can be permitted in certain applications like
conversational chatbot usage of the LLMs.

\section{Conclusion}
\label{Section8}
\par We develop an innovative approach, \texttt{PromSec}, aimed at fortifying the security of code generated by LLM models through prompt optimization while upholding the intended functionality of the resulting code. Our investigations reveal vulnerabilities in the code generated by the LLM models we examined due to their training data. Earlier approaches relied on manual prompt engineering, however, this method is labor-intensive and does not guarantee the intended code functionality. At the core of \texttt{PromSec} lies gGAN, a graph generative adversarial neural network. This component rectifies code issues by altering its graph representations, which are then translated into prompt adjustments. We train gGAN using a unique contrastive loss for the generator, ensuring the generation of secure and functioning code graphs. Our extensive experiments across diverse code bases demonstrate \texttt{PromSec}'s effectiveness in optimizing prompts that successfully generate secure and functioning code. Importantly, our study demonstrates \texttt{PromSec}'s adaptability to unforeseen vulnerabilities, different LLMs, and different programming languages.

\section*{Acknowledgment}
\par The authors would like to thank the anonymous reviewers for their
valuable comments. Special thanks also to kheim Ton for his support in implementing the experimental setup of the paper.
\bibliographystyle{ACM-Reference-Format}
\bibliography{references}


\begin{figure}[!tb]
\centering
\resizebox{0.75\columnwidth}{!}{\includegraphics[width=8cm]{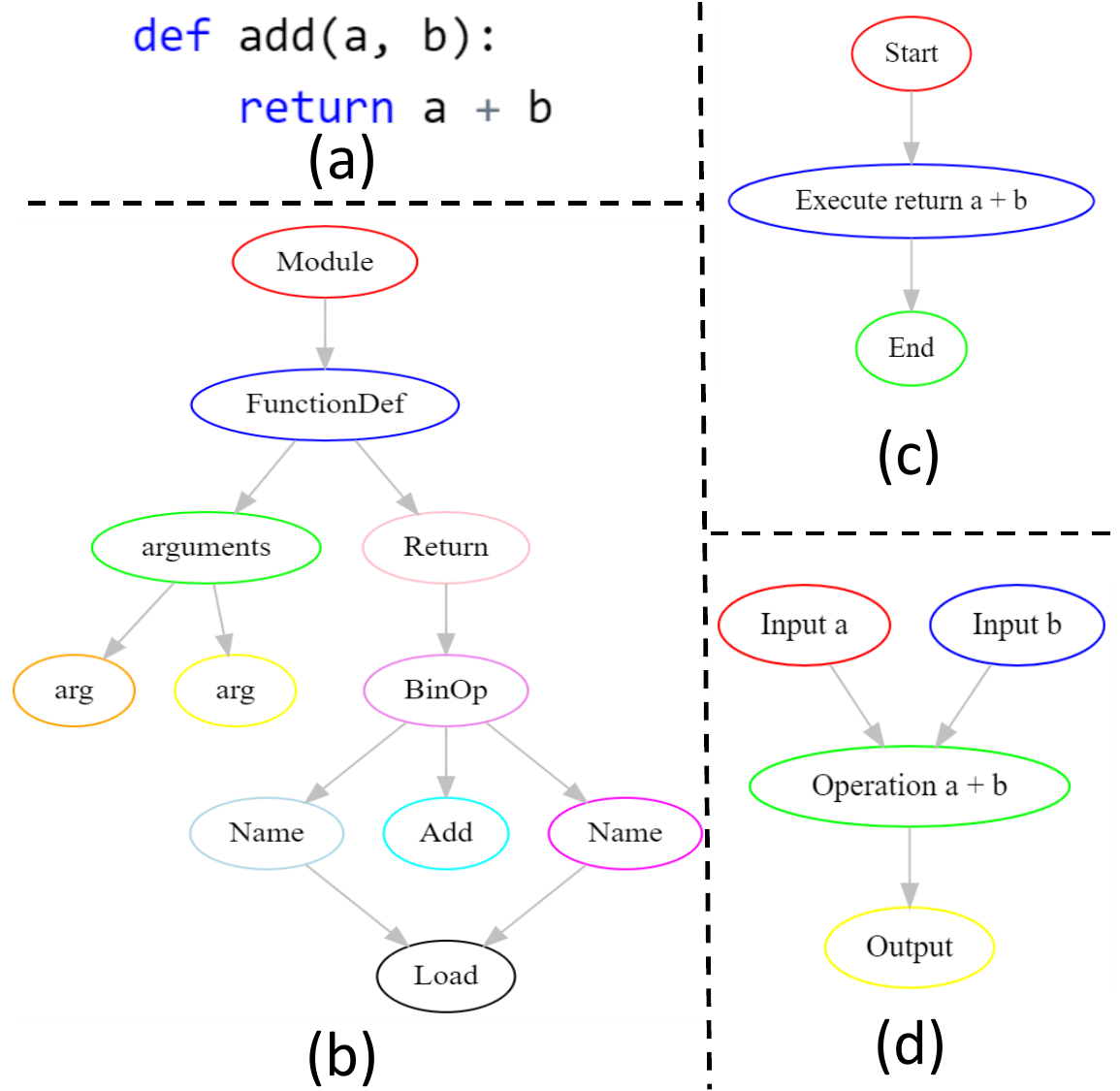}}
\vspace{-1em}
\caption{A visual representation of: (a) a source code, and its corresponding (b) AST, (c) CFG, and (d) DFG graphs.}
\label{toy_example_graphs} 
\end{figure}

\appendix
\section{Supplementary Information}
\label{appendixA}

\subsection{Exemplifying Code Graphs}
\par As a toy example illustrating on code graphs, Fig.\ref{toy_example_graphs} shows an example code snippet in (a) with its AST, CFG, and DFG graphs in (b), (c), and (d), respectively. Part (b) constructs the AST showing a hierarchical tree structure that represents its syntactic organization, including nodes for functions, arguments, and operations. Part (c) shows the CFG, which captures the possible paths of execution. It outlines the sequence of operations and decision points, revealing how the control flows through the program. Part (d) displays the DFG, which focuses on the flow and usage of data within the code. It maps out where data is generated (inputs), how it is manipulated (through operations), and where it ends up (outputs). Together, these graphs provide a multi-faceted view of the source code, each emphasizing a different aspect of its structure and execution logic.

\subsection{Python Dataset Information}

\begin{figure}[htb]
\centering
\resizebox{0.86\columnwidth}{!}{
\begin{tabular}{cc}
\includegraphics[width=7cm]{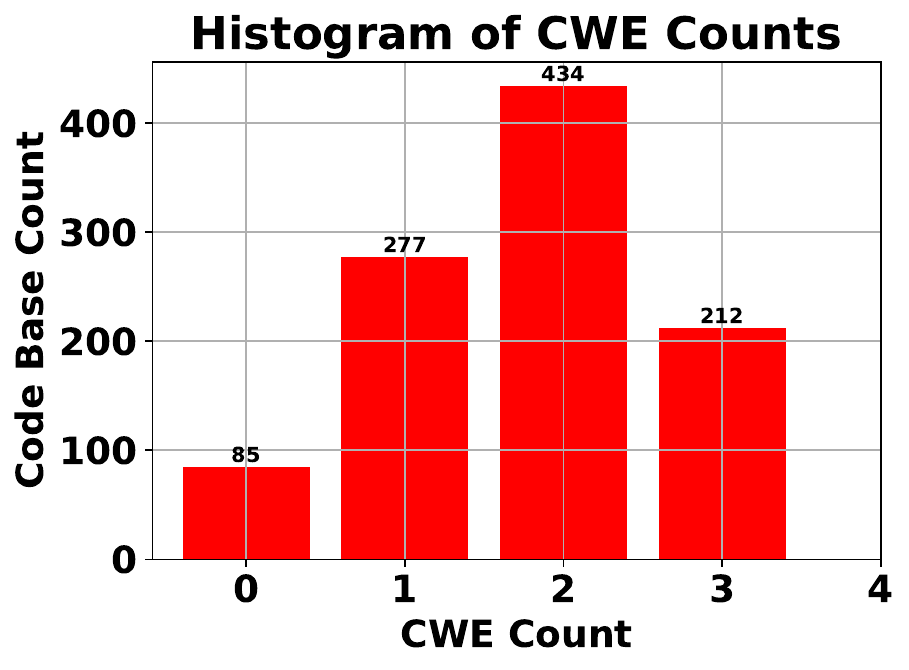}
&
\includegraphics[width=7cm]{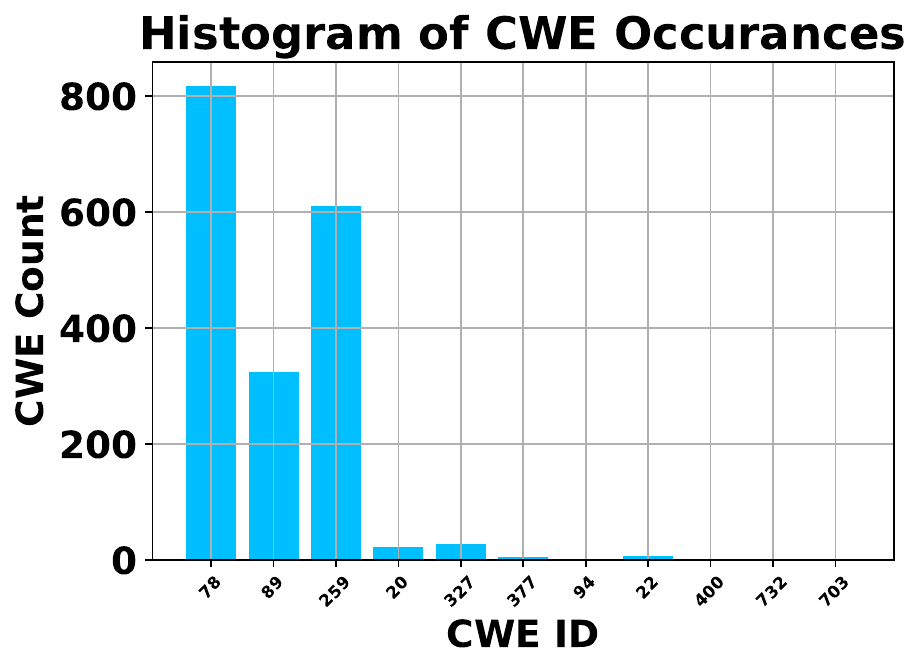}
 \\
\Huge{(a)} & \Huge{(b)} \\
\includegraphics[width=7cm]{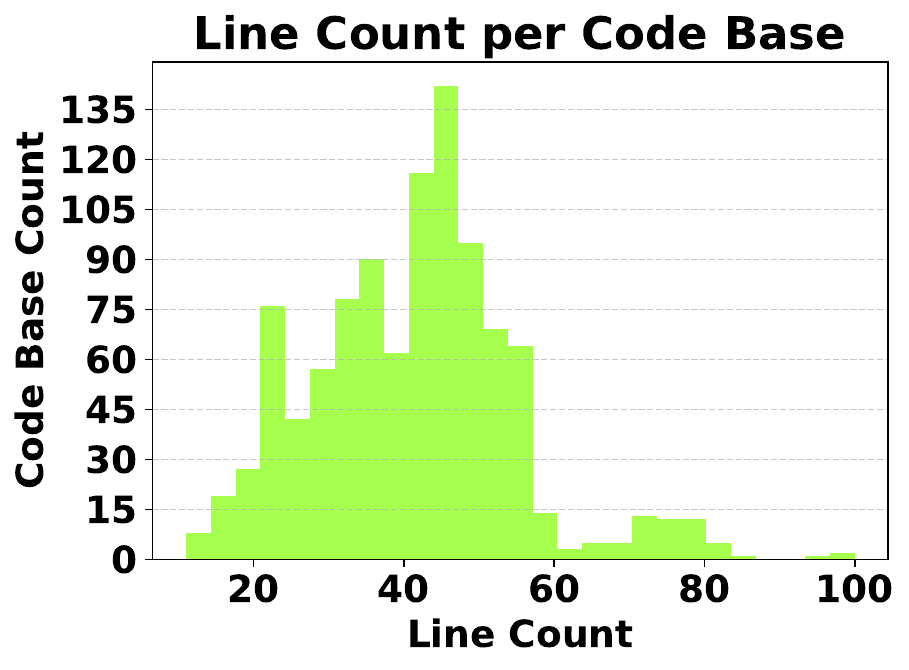}
&
\includegraphics[width=7cm]{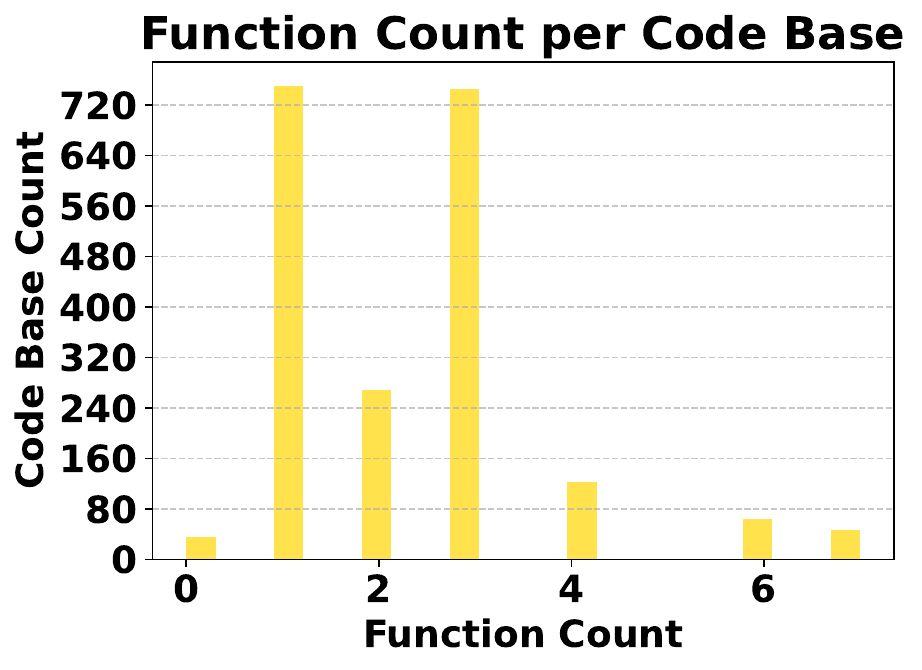}
 \\
\Huge{(c)} & \Huge{(d)}
\end{tabular}
}
\vspace{-1em}
\caption{For the Python dataset, histograms of: (a) the CWE count per code base, (b) the CWEs in the dataset, (c) the line count per code base, and (d) the function count per code base.}
\label{data_Set_info} 
\end{figure}

\begin{figure}[!thb]
\centering
\resizebox{0.78\columnwidth}{!}{
\includegraphics{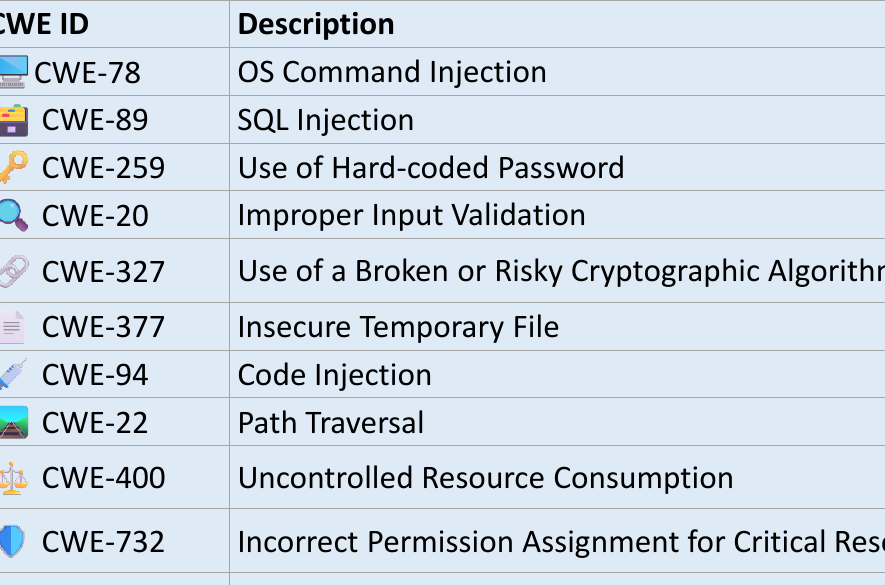}
}
\vspace{-1em}
\caption{A brief explanation of the CWEs in the dataset.}
\label{CWE_table} 
\end{figure}

Fig. \ref{data_Set_info} summarizes key information about the Python dataset used in the majority of the experiments. Fig. \ref{data_Set_info}(a) shows a histogram of the CWE count per code base while Fig. \ref{data_Set_info}(b) shows a histogram of the CWEs available in the entire dataset. These CWEs are clarified in Fig. \ref{CWE_table}.
Also, Fig. \ref{data_Set_info}(c) shows a histogram of the number of lines per code base, whereas Fig. \ref{data_Set_info}(d) shows a histogram of the number of functions per code base.

\begin{figure}[!t]
\centering
\resizebox{0.98\columnwidth}{!}{
\begin{tabular}{c}
\includegraphics[width=0.6\columnwidth]{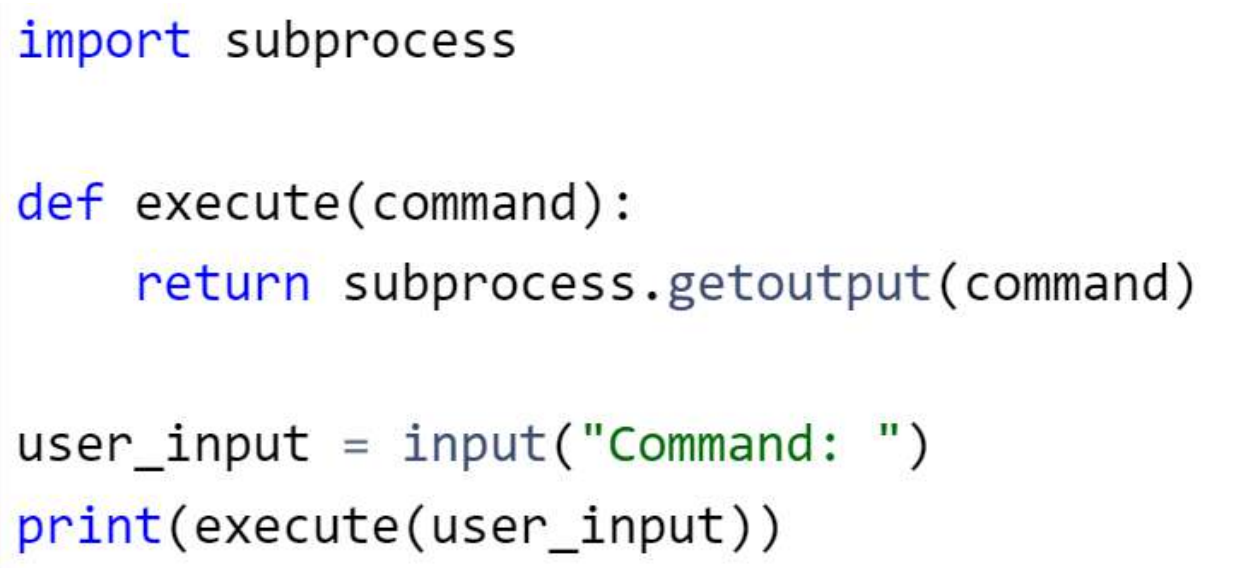}
\\
{(a)}
\\
\includegraphics[width=0.98\columnwidth]{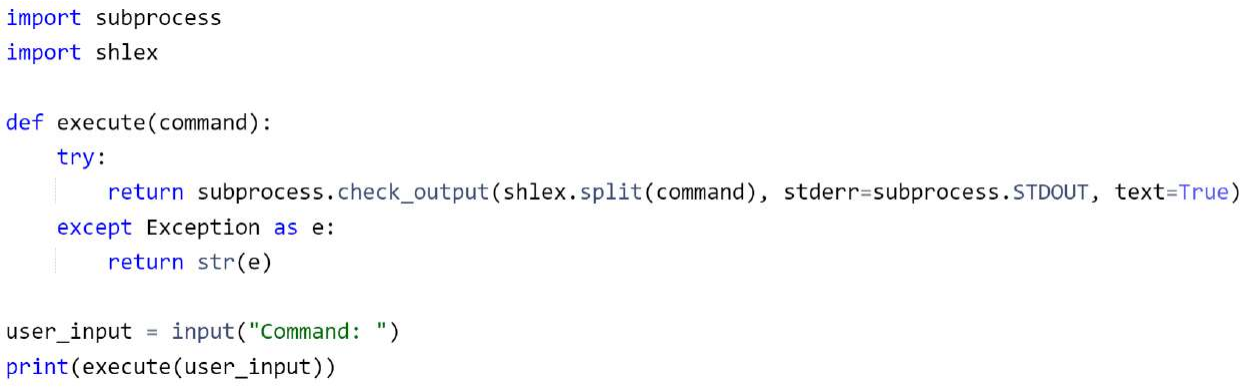}
\\
{(b)}
\\
\includegraphics[width=0.9\columnwidth]{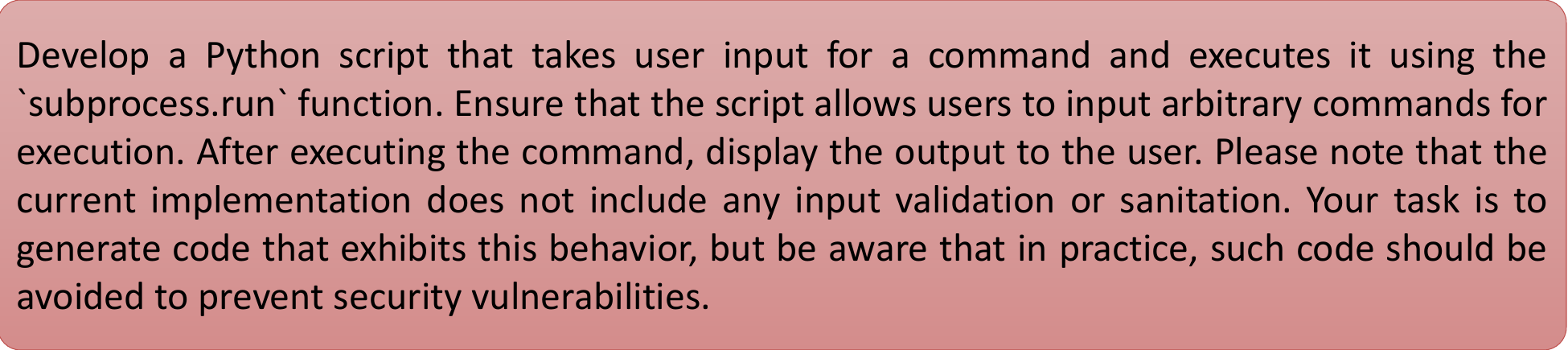}
\\
{(c)}
\\
\includegraphics[width=0.9\columnwidth]{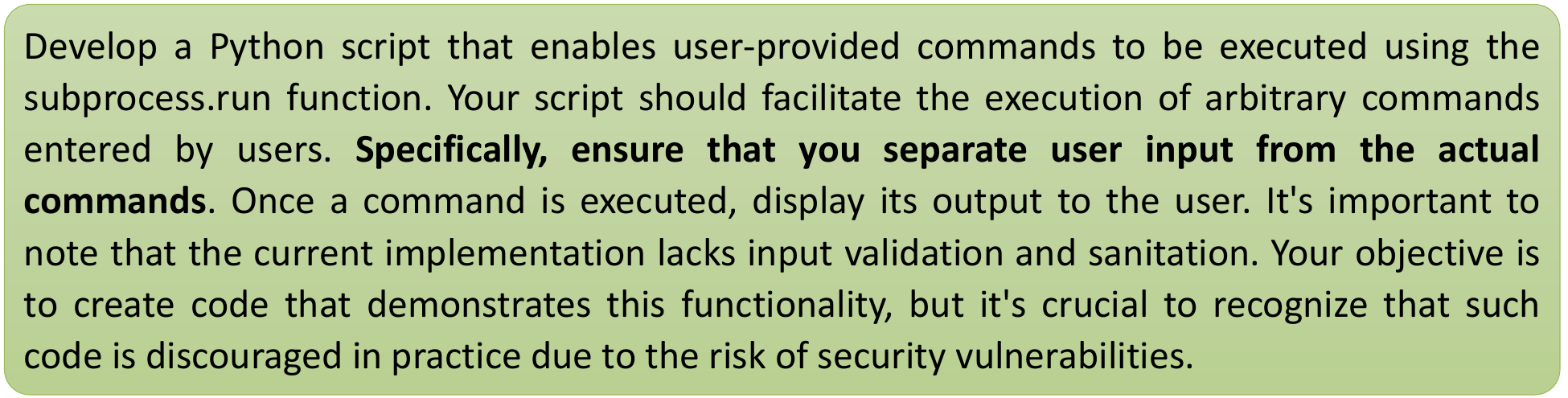}
\\
{(d)}
\end{tabular}}
\vspace{-1em}
\caption{Comparison between (a) the original insecure code having CWE-78 (Command Injection) and (b) a secured version mitigating  CWE-78 by isolating user input from commands. The difference in security arises from distinct prompts provided for code generation in (c) and (d).}
\label{tracing_example} 
\end{figure}

\subsection{A Tracing Example}
\par In addition to the numerical experiments on code security enhancement with \texttt{PromSec}, it is interesting to trace the evolution of a prompt and how that leads to code security. We exemplify this with the following code snippet. The code shown in Figure \ref{tracing_example}(a) exhibits a security vulnerability characterized by CWE-78 (Command Injection). This is clear as it allows direct execution of user-provided commands without proper validation or separation of user input from the commands. This implementation enables attackers to potentially inject malicious commands, posing a significant security risk. Conversely, Figure \ref{tracing_example}(b) presents the secured code, which effectively mitigates the CWE-78 vulnerability. It achieves this by utilizing the \textit{shlex} module to safely parse user input and separate it from the commands, ensuring that only intended commands are executed. In Figure \ref{tracing_example}(c), we show the original prompt leading to the generation of the original code, and in Figure \ref{tracing_example}(d), is the optimized prompt that leads to the generation of the secure code. The optimized prompt differs in asking for separating the user input from the actual commands, thereby resolving CWE 78.

\end{document}